\documentclass[]{aa}
\usepackage{graphicx}
\usepackage{txfonts}
\usepackage{natbib}
\begin{document}
   \title{The nature and origin of Seyfert warm absorbers}

   \author{A. J. Blustin
          \and
          M. J. Page
          \and
          S. V. Fuerst
          \and
          G. Branduardi-Raymont
          \and
          C. E. Ashton
          }
   \offprints{A. J. Blustin\\
             \email{ajb@mssl.ucl.ac.uk}}

   \institute{MSSL, University College London,
             Holmbury St. Mary, Dorking, Surrey RH5 6NT, England
	  }

   \date{Received 2 August 2004 / Accepted 21 October 2004}

   \abstract{We collate the results of recent high resolution X-ray spectroscopic observations of 23 AGN, and use the resulting information to try to provide answers to some of the main open questions about warm absorbers: where do they originate, what effect do they have on their host galaxies, and what is their importance within the energetics and dynamics of the AGN system as a whole? We find that the warm absorbers of nearby Seyferts and certain QSOs are most likely to originate in outflows from the dusty torus, and that the kinetic luminosity of these outflows accounts for well under 1\% of the bolometric luminosities of the AGN. Our analysis supports, however, the view that the relativistic outflows recently observed in two PG quasars have their origin in accretion disc winds, although the energetic importance of these outflows is similar to that of the Seyfert warm absorbers. We find that the observed soft X-ray absorbing ionisation phases fill less than 10\% of the available volume. Finally, we show that the amount of matter processed through an AGN outflow system, over the lifetime of the AGN, is probably large enough to have a significant influence on the evolution of the host galaxy and of the AGN itself.
   }

   \maketitle
%

\section{Introduction}

The warm absorber phenomenon - soft X-ray absorption by ionised gas in our line of sight into the nuclei of certain AGN - was first identified by \citet{halpern1984} in data from an observation of the QSO MR2251-178 with the Einstein observatory. Since then, evidence of warm absorption has been found in about 50\% of nearby Seyfert 1 galaxies using ASCA spectra, and was modelled as consisting of deep photoelectric absorption edges from \ion{O}{vii} and \ion{O}{viii} \citep{reynolds1997,george1998}. Evidence was also found for ionised soft X-ray absorption in BL Lac objects \citep{canizares1984,madejski1991}.

Since the launch of the XMM-Newton and Chandra X-ray observatories, with their high resolution grating spectrometers, our knowledge of the soft X-ray spectra of AGN has been greatly improved. Seyfert warm absorbers are now known to give rise to narrow absorption lines, usually blueshifted by a few hundred km~s$^{\rm -1}$, from elements at a wide range of ionisation levels \citep{kaastra2000,kaspi2000}; these lines are a sensitive diagnostic of the ionisation structure and kinematics of the gas. We now know that the range of ionisation in these absorbers is much wider than previously thought; at the high ionisation end, we see absorption from H-like and He-like iron \citep[e.g.][]{reeves2004}, and the main spectral signature of the lowest ionisation gas is the Unresolved Transition Array (UTA) of M-shell iron which appears at around 16~\AA\, \citep{sako2001,behar2001}. High resolution X-ray spectroscopy of BL Lac objects has not, however, uncovered any evidence for intrinsic ionised absorption \citep{blustin2004bllacs,perlman2004}.

In the light of all of this new observational evidence, it is a good time to re-evaluate our understanding of the phenomenology and significance of warm absorbers. There are two levels of questions which can be asked. Firstly, about the detailed phenomenology of warm absorbers: their ionisation levels, ouflow speeds, and column densities - and how these factors depend on the type and luminosity of the AGN. Secondly, the big questions: where do warm absorbers come from, where do they go, and are they important? What r\^{o}le do they play in the energetics of an active galaxy, and do they tell us something truly fundamental about the way an AGN works and the way it evolves? In this paper, we use the published warm absorber models for fourteen Seyfert 1 type AGN to try to answer some of these questions.

\section{Data and analysis}

\begin{figure}
   \centering
   \includegraphics[width=8.5cm]{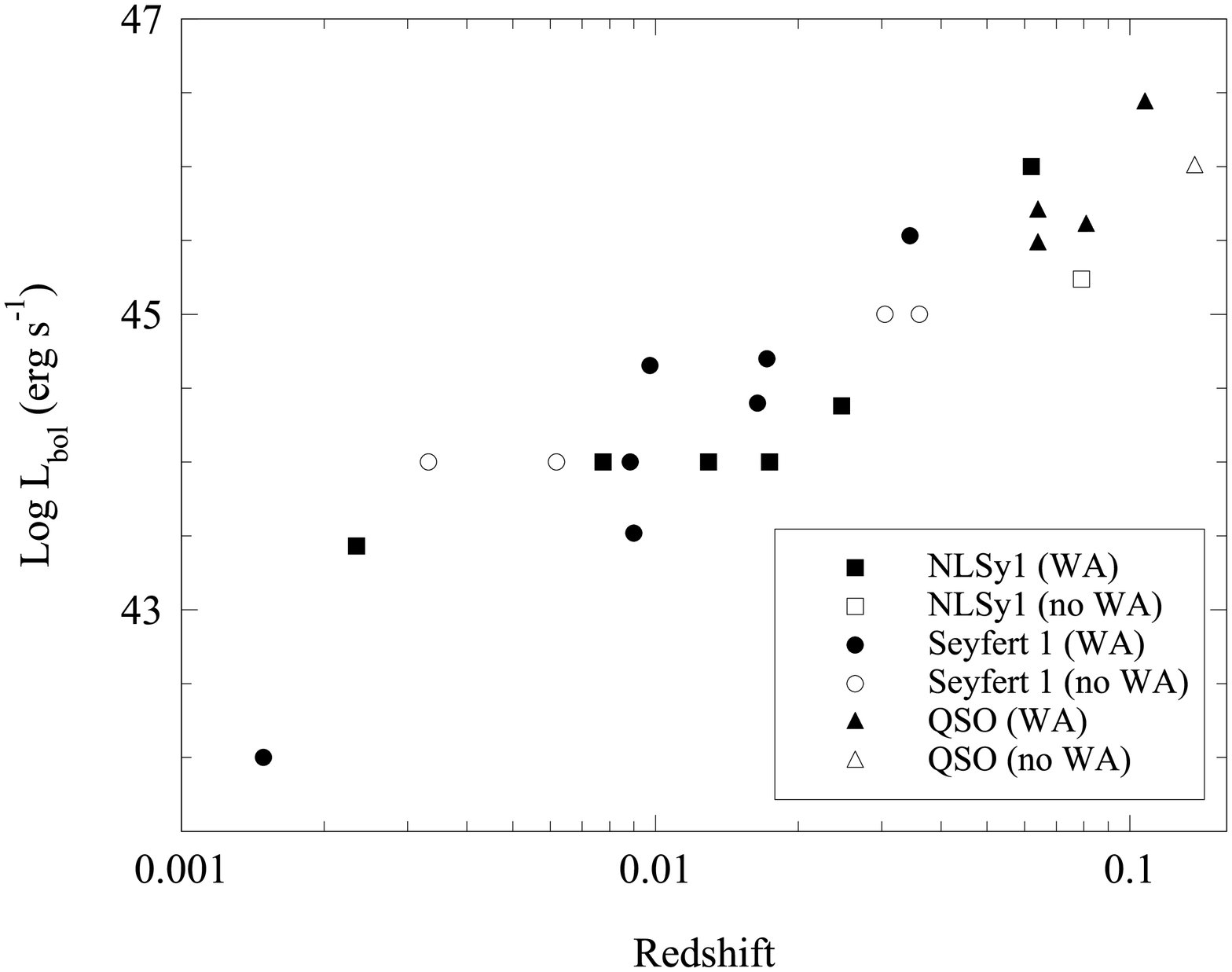}
      \caption{The distribution in redshift and bolometric luminosity of Seyfert 1 type AGN observed with the high-resolution spectrometers on XMM-Newton and Chandra: filled shapes indicate that the presence of a warm absorber has been reported.}
         \label{z_lbol}
   \end{figure}

At the time of writing, high resolution soft X-ray spectra from the grating spectrometers on XMM-Newton and Chandra have been published for 23 Seyfert 1 type AGN (Seyfert 1.x galaxies, Narrow Line Seyfert 1s and Seyfert 1 type quasars). 17 of these objects show evidence of intrinsic ionised absorption, and sufficiently detailed spectral modelling has been applied to the warm absorbers of 14 objects for them to be usable for this study. Table~\ref{all_gals} gives some basic information on all 23 objects, and Fig.~\ref{z_lbol} shows the distribution of the sources in redshift and bolometric luminosity, and whether or not a warm absorber has been detected in each individual source.

   \begin{table*}
    
      \caption[]{Basic properties of Seyfert-type warm absorbers observed with XMM-Newton and Chandra: object name, morphological type, redshift, log L$_{\rm bol}$ (erg s$^{\rm -1}$; with reference if different to that given in the last column), whether the X-ray spectrum is dominated by emission lines (EM), whether the source has a warm absorber (WA), whether a detailed spectral model has been published (Mod), whether the warm absorber is outflowing (Outflow), and reference for the warm absorber model.}
         \label{all_gals}  
   \centering
         \begin{tabular}{llllllllr}
            \hline
            \noalign{\smallskip}
Object     & Type    & z & log L$_{\rm bol}$ & EM & WA & Mod & Outflow & Reference \\
            \noalign{\smallskip}
            \hline
            \noalign{\smallskip}
\object{MR2251-178}      & Sy1 RQQ       & 0.06398 & 45.7$^{\mathrm{a}}$ &   & $\bullet$  & $\bullet$   & $\bullet$       & \citealt{kaspi2003mr}  \\
\object{PG0844+349}      & Sy1 RQQ       & 0.064   & 45.5                &   & $\bullet$ & $\bullet$ & $\bullet$      & \citealt{pounds20030844}  \\
\object{PG1211+143}      & NLSy1 RQQ$^{\mathrm{p}}$ & 0.0809  & 45.6                &   & $\bullet$  & $\bullet$  & $\bullet$      & \citealt{pounds20031211}  \\
\object{PKS 0558-504}    & NLSy1 RLQ     & 0.137   & $\sim$ 46           &   &   &    &        & \citealt{obrien20010558}  \\
\object{IRAS 13349+2438} & NLSy1 RQQ & 0.10764 & 46.4$^{\mathrm{b}}$ &   & $\bullet$ & $\bullet$   & $\bullet$      & \citealt{sako2001}  \\
\object{ESO 141-G55}     & Sy1           & 0.036   & $\sim$ 45           &   &   &    &        & \citealt{gondoin2003}  \\
\object{NGC 4593}        & Sy1           & 0.0084  & 43.5$^{\mathrm{c}}$ &   & $\bullet$ & $\bullet$  & $\bullet$      & \citealt{steenbrugge2003}  \\
\object{NGC 3783}        & Sy1           & 0.00973 & 44.7$^{\mathrm{d}}$ &   & $\bullet$  & $\bullet$  & $\bullet$      & \citealt{netzer2003}  \\
\object{Markarian 509}   & Sy1.2         & 0.0344  & $\geq$ 45.5$^{\mathrm{e}}$ &   & $\bullet$  & $\bullet$  & $\bullet$  & \citealt{yaqoob2003} \\
\object{NGC 7469}        & Sy1.2         & 0.0164  & 44.4$^{\mathrm{f}}$ &   & $\bullet$ & $\bullet$   & $\bullet$     & \citealt{blustin2003}  \\
\object{Markarian 279}   & Sy1.5         & 0.0305  & $\sim$ 45$^{\mathrm{g}}$ &   &   &    &        & \citealt{scott2004}  \\
\object{NGC 3516}        & Sy1.5         & 0.009   & $\sim$ 44           &   & $\bullet$  & $\bullet$   & $\bullet$      & \citealt{netzer2002}  \\
\object{NGC 5548}        & Sy1.5         & 0.01676 & 44.7$^{\mathrm{h}}$ &   & $\bullet$ & $\bullet$  & $\bullet$      & \citealt{steenbrugge20035548}  \\
\object{NGC 4151}        & Sy1.5         & 0.00332 & $\sim$ 44$^{\mathrm{i}}$ & $\bullet$  &   &    &        & \citealt{schurch2004}  \\
\object{NGC 5506}       & Sy1.9 & 0.00618 & $\sim$ 44           & $\bullet$ &           &           &           & \citealt{bianchi2003}  \\
\object{NGC 4258}        & Sy1.9 LINER   & 0.00149 & $\sim$ 42$^{\mathrm{j}}$ & $\bullet$ & $\bullet$ &    &        & \citealt{young2004}  \\
\object{Markarian 478}   & NLSy1 & 0.079$^{\mathrm{k}}$ & 45.2$^{\mathrm{k}}$ &   &   &  &  & \citealt{marshall2003}  \\
\object{Ton S180}        & NLSy1 & 0.06198 & $\sim$ 46$^{\mathrm{l}}$ &   & $\bullet$  &    &        & \citealt{rosanska2004}  \\
\object{MCG -6-30-15}    & NLSy1 & 0.00775 & $\sim$ 44$^{\mathrm{m}}$ &  & $\bullet$ & $\bullet$  & $\bullet$      & \citealt{sako2003}  \\
\object{Markarian 766}   & NLSy1 & 0.012929 & $\sim$ 44$^{\mathrm{n}}$ &   & $\bullet$ & $\bullet$ & & \citealt{mason2003} \\
\object{NGC 4051}        & NLSy1 & 0.00234 & 43.4                &   & $\bullet$ & $\bullet$  &   $\bullet$     & \citealt{ogle2004}  \\
\object{Markarian 359}   & NLSy1 & 0.0174 & $\sim$ 44            &           & $\bullet$ &           & ?         & \citealt{obrien2001359}  \\
\object{Ark 564}         & NLSy1 & 0.02468 & 44.4$^{\mathrm{o}}$ &           & $\bullet$ & $\bullet$ & $\bullet$ & \citealt{matsumoto2004}  \\
            \noalign{\smallskip}
            \hline
         \end{tabular}        
\begin{list}{}{}
\item[$^{\mathrm{a}}$] \citealt{monier2001}; $^{\mathrm{b}}$ \citealt{beichman1986}; $^{\mathrm{c}}$ \citealt{santos1995}; $^{\mathrm{d}}$ \citealt{markowitz2003}; $^{\mathrm{e}}$ \citealt{kriss2000509}; $^{\mathrm{f}}$ \citealt{petrucci2004}; $^{\mathrm{g}}$ \citealt{bachev2003}; $^{\mathrm{h}}$ \citealt{pounds20035548}; $^{\mathrm{i}}$ \citealt{swain2003}; $^{\mathrm{j}}$ \citealt{yuan2002}; $^{\mathrm{k}}$ \citealt{gondhalekar1994}; $^{\mathrm{l}}$ \citealt{turner2002}; $^{\mathrm{m}}$ \citealt{reynolds2000}; $^{\mathrm{n}}$ \citealt{pounds2003766}; $^{\mathrm{o}}$ \citealt{romano2002}; $^{\mathrm{p}}$ \citealt{kaspi2000rblr}
\end{list}
  \end{table*}

Where a source has more than one published detailed warm absorber model, and all of these models provide the values needed for our analysis, we use the most recent model or that which was developed using the highest quality dataset. 

For each of the 14 sources with well modelled warm absorbers, we list the absorbing column, ionisation parameter and outflow velocity of each ionisation phase in Table~\ref{phase_properties}. We use the $\xi$ form of the ionisation parameter, where
   \begin{equation}
      \xi = \frac{L}{n r^{\mathrm{2}}} 
   \end{equation}
in which $L$ is the 1$-$1000~Rydberg source luminosity (in erg s$^{\rm -1}$), $n$ the gas density (in cm$^{\rm -3}$) and $r$ the source distance in cm, so $\xi$ has the units erg cm s$^{\rm -1}$ \citep{tarter1969}. The ionisation parameter can also be expressed in terms of U \citep[see~e.g.][]{netzer1996};
   \begin{equation}
      U = \int_\nu \frac{L_\nu / h\nu}{4 {\pi} n r^{2} c} d\nu \,,
   \end{equation}
where $L_\nu$ is the source luminosity as a function of frequency $\nu$, and the expression can be integrated over the entire Lyman continuum (generally taken as 1$-$1000~Ryd, or $\sim$ 13.6~eV$-$13.6~keV), over the 0.1$-$10~keV range where most of the actual X-ray absorption takes place (U$_{\rm x}$), or over the energy range most relevant for oxygen absorption, namely 0.54$-$10~keV (the `oxygen ionisation parameter' U$_{\rm ox}$). In cases where the authors have given the ionisation parameters of their warm absorber phases in terms of any of these forms of U, we have converted the values to $\xi$ using the relation 
   \begin{equation}
      \xi = U \frac{L h c}{D_{L}^{2} \int_\nu \frac{f_\nu}{\nu} d\nu} 
   \end{equation}
where $L$ is the 1$-$1000~Rydberg source luminosity (in erg s$^{\rm -1}$) as before, $D_L$ is the luminosity distance from us to the source (in cm), and $f_\nu$ is the source flux as a function of frequency. We integrated $f_\nu$ over the Spectral Energy Distribution given by the authors in each case.

   \begin{table}
    
      \caption[]{Equivalent hydrogen columns (log N$_{\rm H}$; cm$^{\rm -2}$), ionisation parameters (log $\xi$; erg cm s$^{\rm -1}$) and outflow velocities (v$_{\rm out}$; km s$^{\rm -1}$) of the modelled warm absorber phases of objects listed in Table~\ref{all_gals}.}
         \label{phase_properties}  
   \centering
         \begin{tabular}{lllr}
            \hline
            \noalign{\smallskip}
Object     & log N$_{\rm H}$ & log $\xi$ & v$_{\rm out}$  \\
            \noalign{\smallskip}
            \hline
            \noalign{\smallskip}
MR2251-178      & 21.51 & 2.9  & 250$^{\mathrm{a}}$ \\
                & 20.3  & 0.68 & 250$^{\mathrm{a}}$ \\
PG0844+349      & 23.6  & 3.7  & 63000 \\
PG1211+143      & 23.7  & 3.4  & 24000 \\
                & 21.8  & 1.7  & 24000 \\
                & 22.9  & -0.9 & not given \\
IRAS 13349+2438 & 22.3  & 2.25 & 0 \\
                & 21.25 & 0    & 420 \\
NGC 4593        & 21.2  & 2.61 & 400 \\
                & 19.8  & 0.5  & 380 \\
NGC 3783        & 21.9  & 1.1  & 750$^{\mathrm{b}}$ \\
                & 22.0  & 2.3  & 750$^{\mathrm{b}}$ \\
                & 22.3  & 2.9  & 750$^{\mathrm{b}}$ \\
Markarian 509   & 21.3  & 1.76 & 200 \\
NGC 7469        & 20.6  & 2.1  & 800  \\
NGC 3516        & 21.9  & 0.78 & 200$^{\mathrm{a}}$ \\
NGC 5548        & 21.68 & 2.69 & 311 \\
                & 21.52 & 1.98 & 440 \\
                & 20.15 & 0.4  & 290 \\
MCG -6-30-15    & 21.3  & 1.25 & 150  \\
                & 21.3  & 2.5  & 1900 \\
Markarian 766   & 21.2  & 0.7  & 0   \\
NGC 4051        & 21    & 1.4  & 402 \\
Ark 564         & 21    & 0    & 150$^{\mathrm{a}}$  \\
                & 21    & 2    & 150$^{\mathrm{a}}$  \\
            \noalign{\smallskip}
            \hline
         \end{tabular}        
\begin{list}{}{}
\item[$^{\mathrm{a}}$] Estimated from the blueshift of the UV absorber
\item[$^{\mathrm{b}}$] Average of two velocity phases at 1000 and 500 km s$^{\rm -1}$
\end{list}
  \end{table}

For each object, we also calculate `averaged' warm absorber properties; Table~\ref{average_properties} lists log weighted average $\xi$ and average outflow speed (both weighted by the absorbing column of each phase), the log total equivalent hydrogen column of the absorber, and the range in log $\xi$ of the absorber phases (estimated here as the difference between log $\xi$ of the highest and lowest modelled ionisation phases).

   \begin{table}
    
      \caption[]{Averaged warm absorber parameters for the sources in Table~\ref{phase_properties}: object name, range in log ionisation parameter ($\Delta$ log $\xi$; erg s$^{\rm -1}$), log total equivalent hydrogen column (log N$_{\rm Htot}$; cm$^{\rm -2}$), log weighted average ionisation parameter (log $\xi$$_{\rm avg}$; erg s$^{\rm -1}$), weighted average outflow velocity (v$_{\rm avg}$; km s$^{\rm -1}$).}
         \label{average_properties}  
   \centering
         \begin{tabular}{llllr}
            \hline
            \noalign{\smallskip}
Object     & $\Delta$ log $\xi$ & log N$_{\rm Htot}$ & log $\xi$$_{\rm avg}$ & v$_{\rm avg}$  \\
            \noalign{\smallskip}
            \hline
            \noalign{\smallskip}
MR2251-178      & 2.22    & 21.54 & 2.87 & 250 \\
PG0844+349      & 0       & 23.6  & 3.7  & 63000 \\
PG1211+143      & 4.3     & 23.77 & 3.39 & 24000 \\
IRAS 13349+2438 & 2.25    & 22.34 & 2.21 & 30 \\
NGC 4593        & 2.11    & 21.82 & 2.54 & 204 \\
NGC 3783        & 1.8     & 22.58 & 2.68 & 750 \\
Markarian 509   & 0       & 21.3  & 1.76 & 200 \\
NGC 7469        & 4       & 20.60 & 2.1  & 800 \\
NGC 3516        & 0       & 21.9  & 0.78 & 200 \\
NGC 5548        & 2.29    & 21.92 & 2.51 & 362 \\
MCG -6-30-15    & 1.25    & 21.60 & 2.22 & 1025 \\
Markarian 766   & unknown & 21.2  & 0.7  & 0   \\
NGC 4051        & no info & 21    & 1.4  & 402 \\
Ark 564         & 2       & 21.30 & 1.70 & 150 \\
            \noalign{\smallskip}
            \hline
         \end{tabular}        
  \end{table}

\section{Results and discussion}

\subsection{The prevalence of warm absorbers}

A fundamental question about warm absorbers is whether their presence is somehow essential to an AGN - or put another way, are there Seyferts which do not have them? Previous work (e.g. \citealt{reynolds1997}) found that only about half of AGN have warm absorbers (or at least warm absorbers detectable with instrumentation prior to XMM-Newton and Chandra). The group of AGN we examine here is not an ideal sample to answer this question, because many of them were selected for observation precisely because they were known to exhibit ionised soft X-ray absorption from previous studies. There are, however, certain objects in the group which do not show evidence of warm absorption.

Of a total of 23 objects in Table~\ref{all_gals}, eleven are ordinary Seyferts (as classified by the authors referenced in the table, or otherwise by NED), seven are NLSy1s, and five are quasars. Four of the Seyferts lack a warm absorber. Two of these, NGC 4151 and NGC 5506, are emission line sources with an obscured central engine. The presence of the narrow X-ray emission lines implies that NGC 4151, at least, would appear to have a warm absorber if viewed from a different angle. The Chandra observation of Markarian 279 \citep{scott2004} was very short, and the source was in a low flux state, so the signal-to-noise of the spectrum is poor. Although \citet{scott2004} found no significant evidence of X-ray warm absorption, this source does have an intrinsic UV absorber, making it very likely that an X-ray absorber will be found in better quality data in future; \citet{crenshaw1999} found that objects in their sample with UV absorption always showed evidence of an X-ray absorber. The one remaining object without a warm absorber, ESO 141-G55 \citep{gondoin2003} is towards the high end of the L$_{\rm bol}$$-$redshift distribution, but there does not seem to be anything significantly unusual about it with respect to the other sources. NGC 4258 is basically a Seyfert 2-type object, without a soft X-ray absorber; it has however been claimed to exhibit absorption lines above 6~keV \citep{young2004}.

Of the NLSy1s, only Markarian 478 lacks a warm absorber. Like ESO 141-G55, it has a fairly high luminosity, but does not seem to have very different properties from other sources which do have warm absorbers. One of the quasars (PKS 0558-504, \citealt{obrien20010558}) does not show evidence of warm absorption. It is a radio loud quasar, though, and radio-loud objects might be particularly problematic in terms of finding observational evidence for ionised outflows (see, for example, the non-detection of intrinsic ionised absorption in BL Lac objects; \citealt{blustin2004bllacs,perlman2004}). Disregarding the emission line and radio loud objects in the sample, as well as Markarian 279 since it has a UV absorber, we find that 2 out of 18 Seyfert 1-type objects in the sample (11\%) show no evidence of an ionised outflow. If this was representative of the population as a whole, it would imply a global covering factor of almost 90\% for X-ray warm absorbers.

It is as well to be careful of reading too much into the non-detection of warm absorption in ESO 141-G55 and Markarian 478. An absorber might be present, but not along our line of sight, or with a turbulent velocity and column density too low to be observed with today's instrumentation; orientation effects could be very important in determining the appearance of AGN outflows \citep[c.f.][]{elvis2000}. If indeed there are AGN which genuinely lack warm absorbers, then we would know that AGN were able to do without them and they must be a side-effect - rather than an essential feature - of the AGN process.

\subsection{Phenomenology of warm absorbers}

As we discussed in Section 2, the emerging picture of warm absorbers is that they usually have multiple ionisation phases and are flowing away from the central engine at speeds of a few hundred km s$^{\rm -1}$. Within this general conception, are any particular patterns apparent?

As far as the multi-phase ionisation structure is concerned, there is currently no consensus as to whether this takes the form of discrete ionisation phases \citep[i.e.][]{krongold2003}, a continuous ionisation parameter distribution \citep[e.g.][]{krolik2001,rosanska2004}, or some combination of these \citep[c.f.][]{blustin2002}. All global spectral modelling of warm absorbers, to date, has assumed that the gas contains discrete ionisation phases; for the 14 objects in our sample, the average number of modelled ionisation phases is two. 

We contend, however, that the number of modelled phases does not tell us anything fundamental about the ionisation structure, as it is more likely to be a function of the spectroscopic knowledge and techniques of the investigator. Indeed, two of the highest statistical quality spectra - those of NGC 3783 \citep{netzer2003} and NGC 5548 \citep{steenbrugge2003} - require three ionisation phases, and it is quite possible that better quality spectra of other objects would also require a larger number of phases. 

Perhaps a more interesting quantity is the range of ionisation parameter in the absorber (as listed in Table~\ref{average_properties}). If all warm absorbers had the same ionisation structure, one might expect that increasing the overall total column N$_{\rm Htot}$ of the absorber would bring a greater range of ionisation states into view. In reality, the objects in our sample show no such correlation, which either implies a wide variation in the ionisation structures, or, very probably, significant differences in the analysis methods and statistical quality of the spectra used by different authors. In particular, observers may be looking mainly for `traditional' warm absorber ions such as \ion{O}{viii} and \ion{O}{vii}, and others at a similar ionisation level, and may not be expecting to find much more lowly ionised species for which atomic data are harder to obtain.

\begin{figure}
   \centering
   \includegraphics[width=8.5cm]{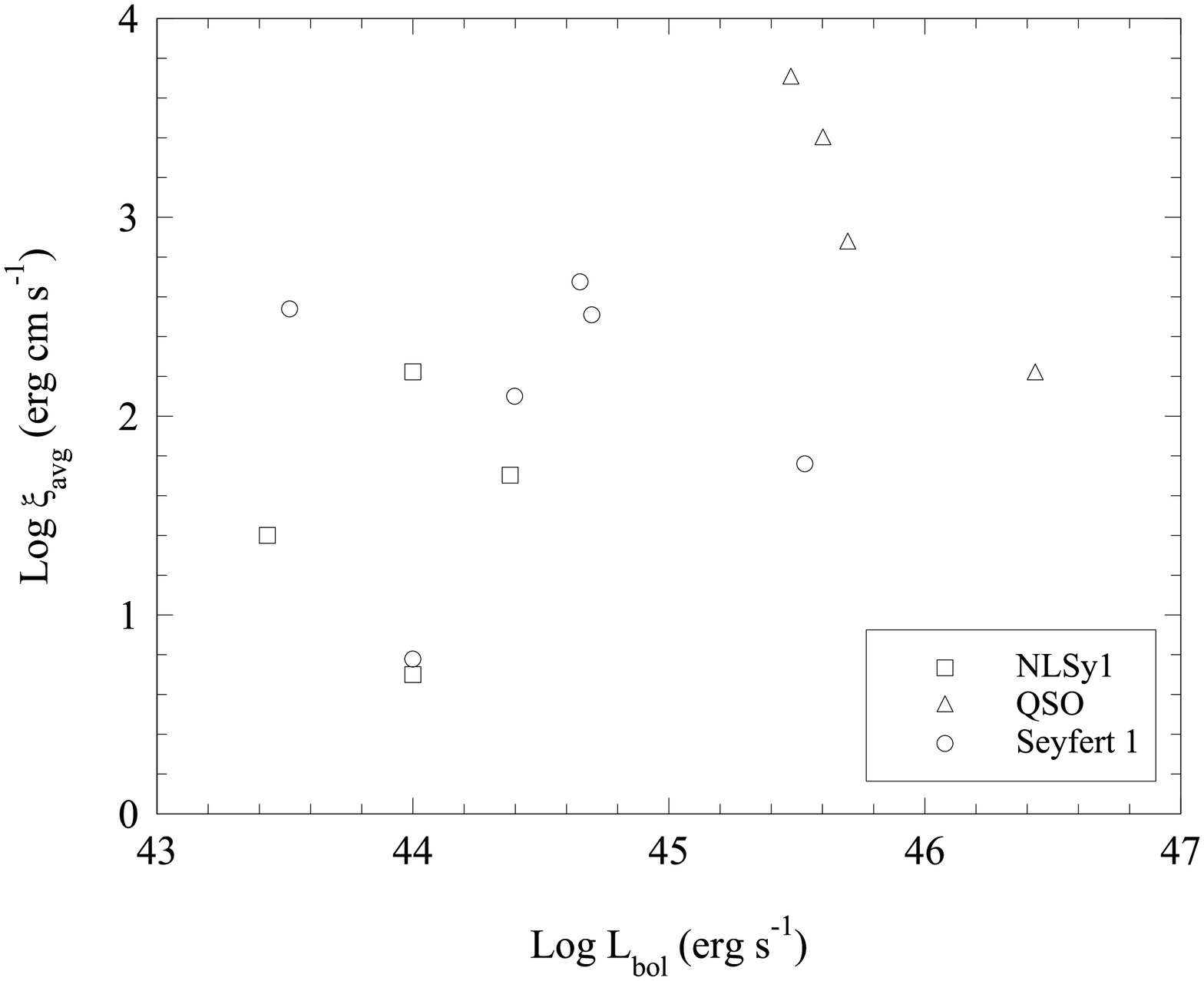}
      \caption{Log $\xi$$_{\rm avg}$ versus log L$_{\rm bol}$ for the objects in Table~\ref{average_properties}.}
         \label{lbol_xiavg}
   \end{figure}

The average ionisation parameter of an absorber, though a crude measure, is probably less open to interpretation than its ionisation range, and may be more useful for comparative purposes. The distance scales of an AGN environment are expected to be roughly proportional to the bolometric luminosity of the central engine, and so since (according to the definition of the ionisation parameter), $\xi$ decreases faster with distance than it increases with luminosity, one might expect the average ionisation parameter of warm absorbers to decrease with increasing source luminosity. In fact, there is no strong correlation between log average ionisation and log bolometric luminosity for the sources in Table~\ref{average_properties} (linear correlation coefficient C = 0.52, probability p of getting C greater than or equal to this from a random distribution = 0.04). There are slight differences between log $\xi$$_{\rm avg}$ for each class of AGN; for the quasars, log $\xi$$_{\rm avg}$ = 3.0 $\pm$ 0.6, for the Seyferts log $\xi$$_{\rm avg}$ = 2.1 $\pm$ 0.7 and for the NLSy1s log $\xi$$_{\rm avg}$ = 1.5 $\pm$ 0.6, where the errors are given as the standard deviation. The overall log $\xi$$_{\rm avg}$ for all types is 2.2 $\pm$ 0.8. Log $\xi$$_{\rm avg}$ is plotted versus log L$_{\rm bol}$ in Fig.~\ref{lbol_xiavg}.

\begin{figure}
   \centering
   \includegraphics[width=8.5cm]{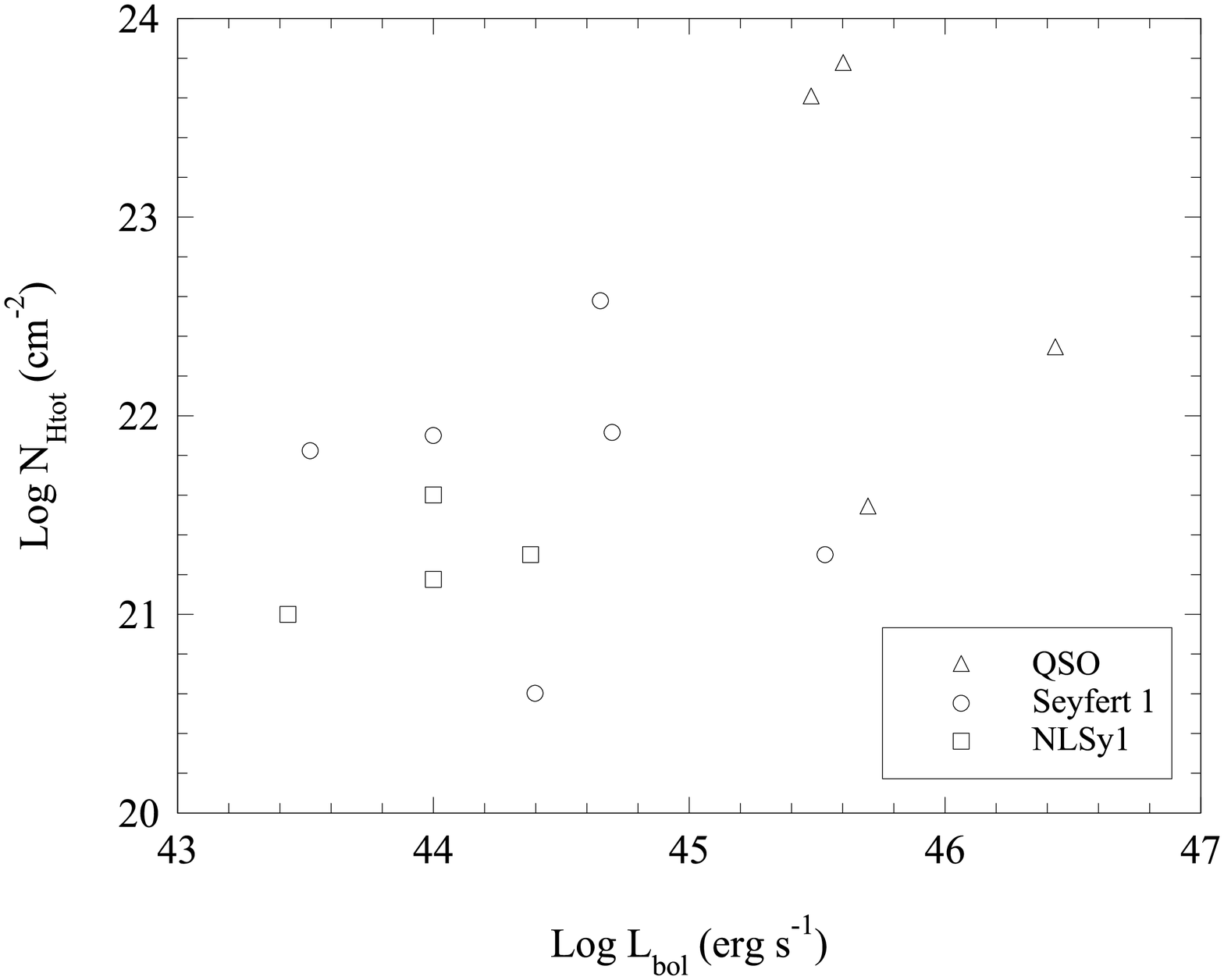}
      \caption{Log N$_{\rm Htot}$ versus log L$_{\rm bol}$ for the objects in Table~\ref{average_properties}.}
         \label{lbol_nhtot}
   \end{figure}

There is also no strong correlation between log total column and log bolometric luminosity (C = 0.50, p = 0.03; see Fig.~\ref{lbol_nhtot}), but again, the three classes of object have slightly different log average columns: 22.8 $\pm$ 0.9, 21.7 $\pm$ 0.6 and 21.3 $\pm$ 0.2 cm$^{\rm -2}$ for quasars, Seyferts and NLSy1s respectively. The log average column for all objects is 21.9 $\pm$ 0.9 cm$^{\rm -2}$. There is, however a significant correlation between log $\xi$$_{\rm avg}$ and log N$_{\rm Htot}$: C = 0.70, p = 0.003. The strength of the correlation decreases to C = 0.44 and p = 0.01 for log $\xi$ and log N$_{\rm H}$ of the individual phases (Table~\ref{phase_properties}). This correlation is simply an observational bias, since the more highly ionised the warm absorber, the greater the column required for it to be observable.

\begin{figure}
   \centering
   \includegraphics[width=8.5cm]{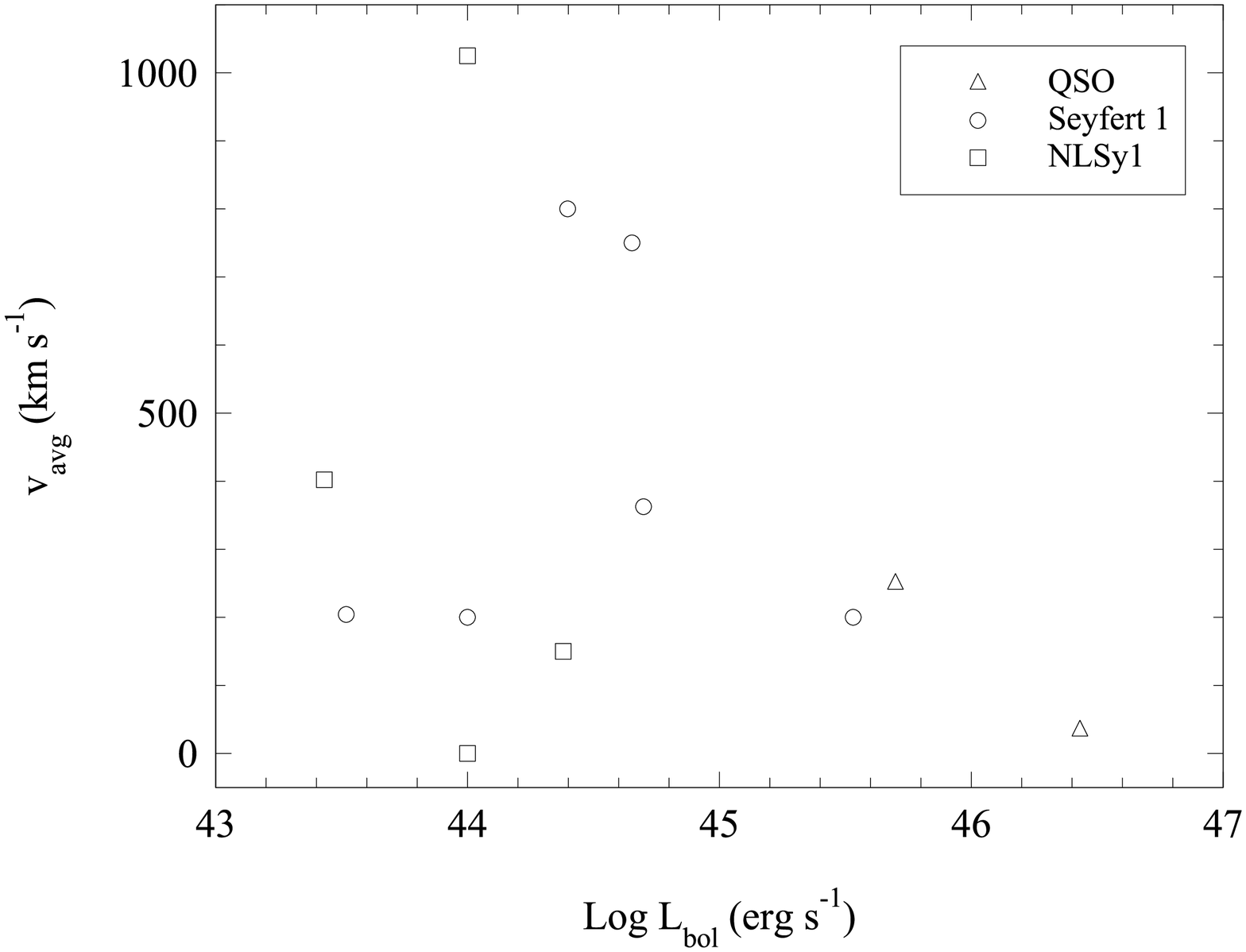}
      \caption{Average outflow velocity versus log L$_{\rm bol}$ for the objects in Table~\ref{average_properties}, excluding PG0844+349 and PG1211+143.}
         \label{lbol_vavg}
   \end{figure}

\begin{figure}
   \centering
   \includegraphics[width=8.5cm]{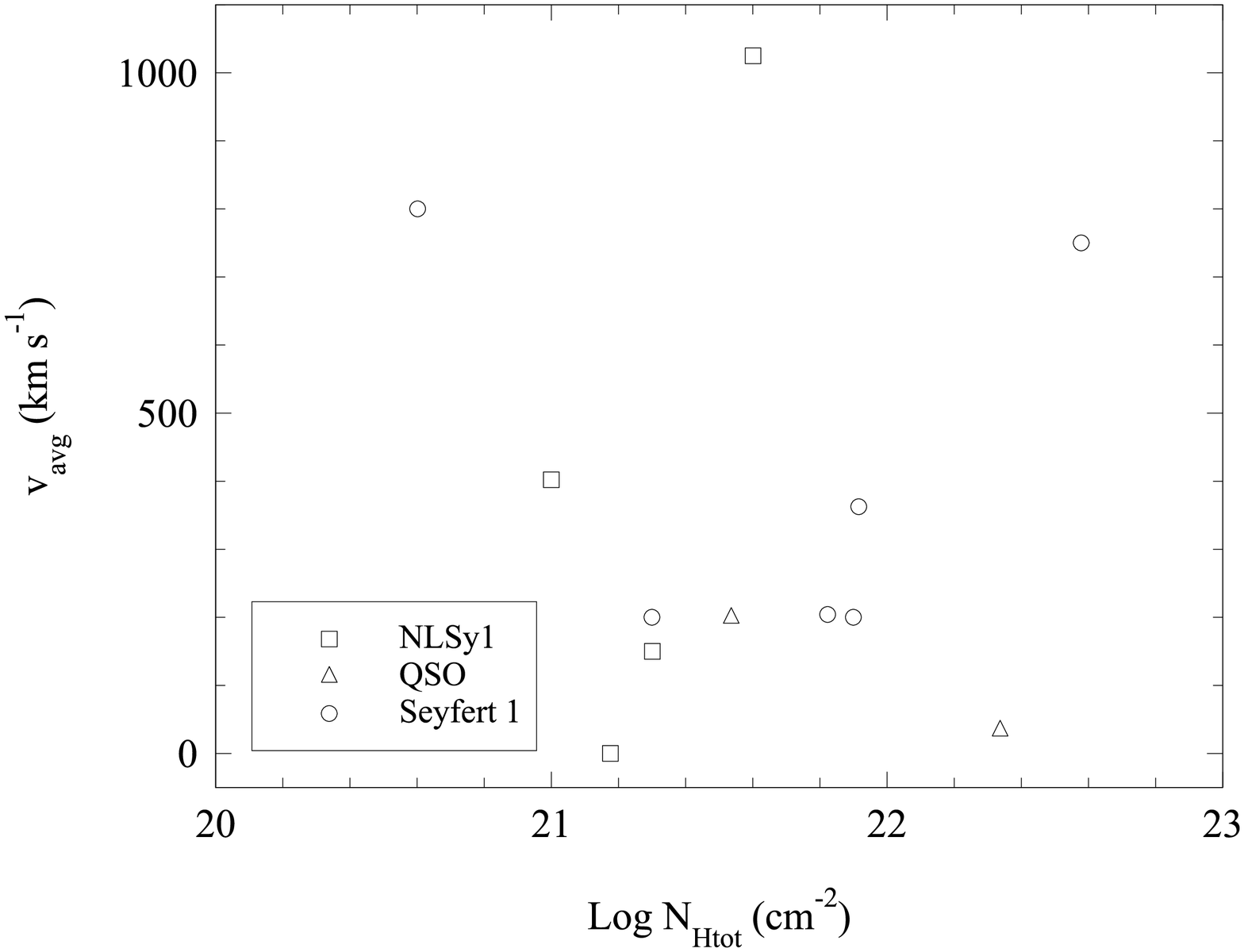}
      \caption{Average outflow velocity versus log N$_{\rm Htot}$ for the objects in Table~\ref{average_properties}, excluding PG0844+349 and PG1211+143.}
         \label{nhtot_vavg}
   \end{figure}

\begin{figure}
   \centering
   \includegraphics[width=8.5cm]{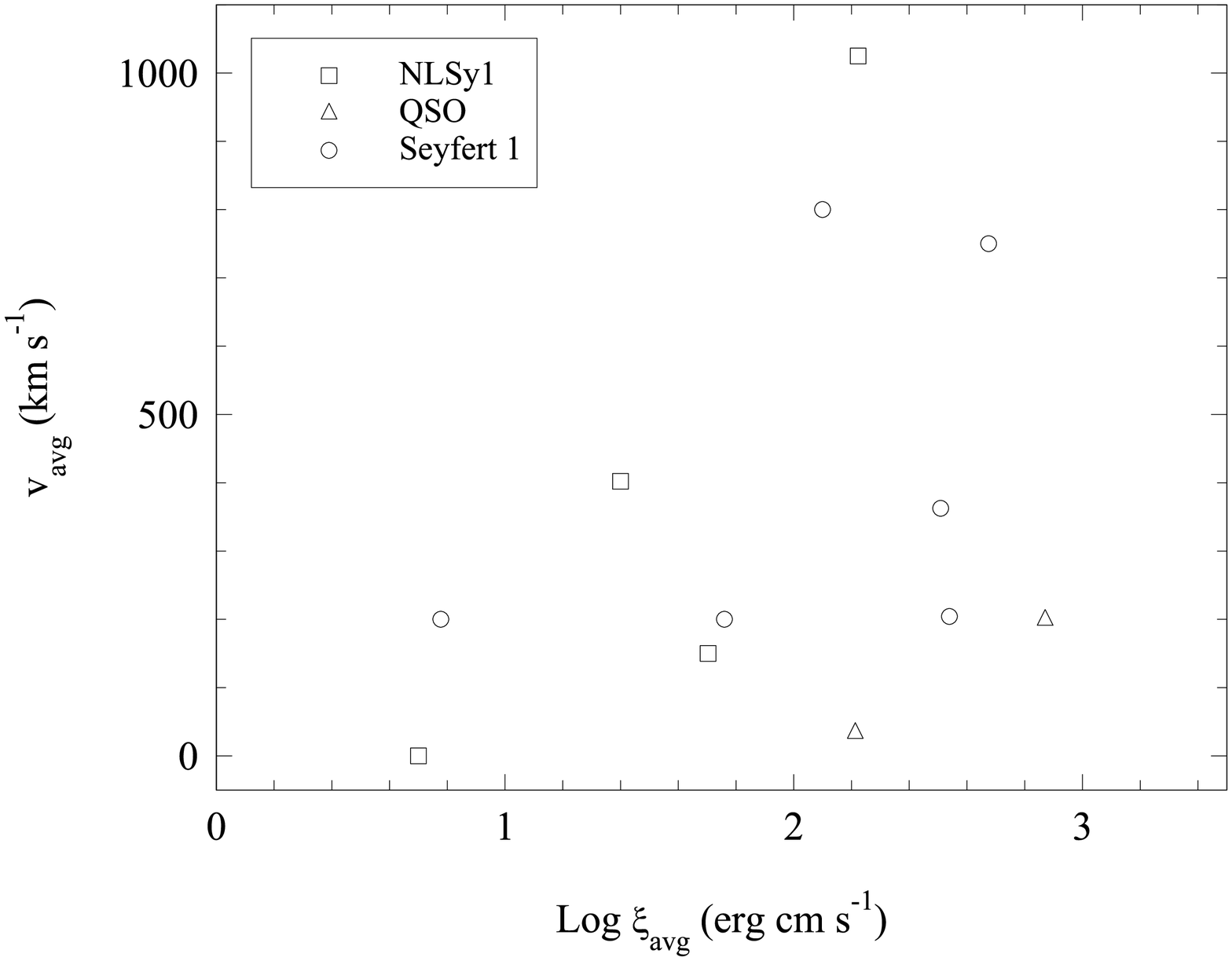}
      \caption{Average outflow velocity versus log $\xi$$_{\rm avg}$ for the objects in Table~\ref{average_properties}, excluding PG0844+349 and PG1211+143.}
         \label{xiavg_vavg}
   \end{figure}

Apart from the ionisation parameter and column, a warm absorber is also characterised by its outflow velocity. It is immediately obvious from Table~\ref{average_properties} that we are dealing with two rather different orders of outflow speed. The Seyferts and NLSy1s have average warm absorber outflow speeds of a few hundred km s$^{\rm -1}$; the two PG quasars have outflow speeds of tens of thousands of km s$^{\rm -1}$ (though we note that \citet{kaspi2004} interprets the RGS absorption lines in PG1211+143 as indicating an outflow speed of only 3000~km s$^{\rm -1}$). If the PG quasars are excluded from the analysis, the average outflow speeds are not significantly correlated with bolometric luminosity (Fig.~\ref{lbol_vavg}), absorbing column (Fig.~\ref{nhtot_vavg}) or average ionisation parameter (Fig.~\ref{xiavg_vavg}). 

In a radiatively accelerated wind, the density of the outflow decreases with increasing velocity, so we expect more highly ionised gas to be outflowing faster. It is therefore, perhaps, surprising that there is no correlation between the average ionisation parameters and velocities. Even the individual phases for each object, although their ionisations may vary, are outflowing at the same or similar speeds. The one exception is MCG -6-30-15 in which the high-ionisation phase is outflowing much faster. 

Realistically, it might be hard to observe an increasing velocity with ionisation parameter due to the rapid fall-off with density in such winds. Also, the two outflow phases in MCG -6-30-15 might not be part of the same accelerating wind - the high velocity, high ionisation component might originate from closer in to the nucleus where the escape velocity is very high, and the low velocity component could be launched from much further away. 

If lower and higher ionisation phases are outflowing with the same speed, this may indicate that they are in pressure equilibrium \citep{krolik2001}. It is difficult to test this directly though as the pressure form of the ionisation parameter, $\Xi$, which is used in studies of thermal equilibrium in outflows, is hard to estimate.
   \begin{equation}
     \Xi = \frac{\xi}{4 \pi c k T} \,,
   \end{equation}
where $c$ is the speed of light, $k$ is Boltzmann's constant and T is the gas temperature \citep{krolik1981}.


The relativistic velocities of the PG quasar outflows are not simply due to the high luminosities of the sources - IRAS 13349+2438 and MR2251-178 are both quasars with even higher bolometric luminosities, but they have low velocity ionised outflows very much like those seen in nearby Seyferts. There has to be some other fundamental difference in the origin of the outflow.

\subsection{Energetics of warm absorbers}

How much mass is carried out of the AGN by the outflow? How does this compare to the amount of matter being accreted? Do ionised outflows carry a significant fraction of the energy output of an AGN? The answers depend upon the way one models the outflow. The most straightforward assumption is that the warm absorber contains a gas with uniform filling factor and density smoothly decreasing with 1/r$^{\rm 2}$. This is unlikely to be a realistic scenario, though; observations \citep[e.g.][]{kinkhabwala2002} have shown that a range of densities and ionisation parameters is likely to exist at any given radius, and therefore that the warm absorber/emitter is more likely to consist of filaments or clouds with a low overall filling factor. We have to bear this in mind when estimating the mass outflow rate and energy output of the AGN.

For a spherical outflow, the mass contained within a segment of a thin spherical shell of radius $r$, thickness $\delta$$r$, solid angle $\Omega$ and mass density $\rho$$(r)$ is

   \begin{equation}
      M = r^2 {\rho}{(r)} {\delta}{r} \Omega  \,. 
   \end{equation}

If matter is moving through this shell at a speed $\delta$$r$/$\delta$$t$ = $v(r)$, then the mass outflow rate through the shell is
   \begin{equation}
      \dot M = r^2 {\rho}{(r)} v(r) \Omega  \,. 
   \end{equation}

Now assuming that the outflow has cosmic elemental abundances (i.e. $\sim$ 75\% by mass of hydrogen and $\sim$ 25\% by mass of helium), $\rho$$(r)$ $\sim$ 1.23$m_p$${\bar n}(r)$ where $m_p$ is the mass of a proton and ${\bar n}(r)$ is the average ion number density at radius $r$, giving
   \begin{equation}
\label{m_out_eqn}
      \dot M \sim 1.23 r^2 m_p {\bar n}(r) v(r) \Omega  \,. 
   \end{equation}

In a continuous spherical outflow at constant velocity, ${\bar n}(r)$ must fall off as 1/r$^{\rm 2}$, giving the relationship
   \begin{equation}
     {\bar n}(r) = \frac{k}{r^2} \,. 
   \end{equation}
where $k$ is a constant.

Let us say that the base of the outflow is at r = R, where 
   \begin{equation}
     {\bar n}(R) = \frac{k}{R^2} \,, 
   \end{equation}
which gives
   \begin{equation}
      k = {\bar n}(R) R^2
   \end{equation}
and so
   \begin{equation}
      {\bar n}(r) = \frac{{\bar n}(R) R^2}{r^2} \,. 
   \end{equation}
Substituting this into Eqn.~\ref{m_out_eqn}, and noting that $v(r)$ is simply $v$ for the constant velocity outflow we are assuming, we get 
   \begin{equation}
      \dot M \sim 1.23 r^2 m_p v \Omega \frac{{\bar n}(R) R^2}{r^2}   
   \end{equation}
and therefore 
   \begin{equation}
      \dot M \sim 1.23 R^2 m_p {\bar n}(R) v \Omega \,.    
   \end{equation}

We need to relate ${\bar n}(R)$ to a measurable quantity. The ionisation parameter is given by
   \begin{equation}
      {\xi} = \frac{L_{ion}}{n(R) R^2} \,.    
   \end{equation}
Now, the density $n(R)$ in the ionisation parameter expression refers to the `microscopic' electron density in the gas where the physical absorption is taking place. We assume here that the electron and ion number densities are similar, i.e. that hydrogen is fully ionised. The density in the mass outflow expression ${\bar n}(R)$ refers to a macroscopic ion number density averaged across the segment of shell. It is therefore a function of both the number density of the gas that is actually absorbing at ionisation parameter $\xi$, and also the volume filling factor of this gas $C_v$$(R)$, so that
   \begin{equation}
      {\bar n}(R) = n(R) {C_v}{(R)} \,. 
   \end{equation}

This then gives us
   \begin{equation}
      \dot M \sim 1.23 R^2 m_p n(R) {C_v}{(R)} v \Omega \,,    
   \end{equation}
and substituting in the expression for the ionisation parameter,
   \begin{equation}
      \dot M \sim 1.23 R^2 m_p {C_v}{(R)} v \Omega \frac{L_{ion}}{{\xi} R^2}  \,,    
   \end{equation}
giving a final expression for the mass outflow rate of 
   \begin{equation}
      \dot M \sim \frac {1.23 m_p L_{ion} {C_v}{(R)} v \Omega}{{\xi}}  \,.    
   \end{equation}

We can then use this expression to estimate the mass outflow rate for the warm absorbers of the 14 AGN in our sample. $\xi$ and $v$ are directly measurable, and are listed in Table~\ref{average_properties}. We estimated L$_{\rm ion}$ for each source by integrating over SEDs based on that used for NGC~5548 by \citet{kaastra20025548}, since this is the SED used in calculating the SPEX 2.00 \citep{kaastra2002spex} warm absorber model which we use below. In these SEDs, the 0.7$-$13.6~keV X-ray power-law (with slopes taken from the references in Table~\ref{all_gals}) joins on to a soft excess approximated by a $\Gamma$ = 3.69 power-law from 0.3$-$0.7~keV, and a $\Gamma$ = 2 power-law from 0.3~keV down to 13.6~eV. The slope of the 0.3$-$0.7~keV power-law was determined by constraining the ratio of F$_{\nu}$ in the power-law and soft excess ($\Gamma$ = 2 component) at 0.7~keV to be the same as that in the NGC~5548 SED. The values of L$_{\rm ion}$ are listed in Table~\ref{phase_wa_energetics}. We estimate the solid angle of the outflow using the information that $\sim$ 25\% of nearby AGN are type 1 \citep{maiolino1995}, and that the covering factor of these outflows seems to be at least 50\% \citep{reynolds1997}. This gives $\Omega$ $\sim$ 1.6. 

The volume filling factor $C_v$ of each observed phase is harder to estimate, especially since it very probably has some radial dependence. We cannot therefore take the same value for each AGN, as their warm absorbers could be at a wide range of radii with different (unknown) morphologies. The fact that $C_v$$(R)$ must be less than or equal to 1 would allow us to calculate an upper limit to the mass outflow rate. We can, however, estimate the volume filling factor since the momentum of the outflow (which is dependent upon $C_v$$(R)$) must be of the order of the momentum of the radiation it absorbs (which we can estimate using the published warm absorber models) plus the momentum of the radiation it scatters (by Thomson scattering, in this approximation, which is estimated using the absorbing column):
   \begin{equation}
      {\dot M} v \sim {\dot P_{abs}} + {\dot P_{scatt}}  \,,    
   \end{equation}
where
   \begin{equation}
      {\dot P_{abs}} = \frac {L_{abs}}{c}  \,,    
   \end{equation}
in which $L_{abs}$ is the luminosity absorbed by the outflow over the whole 1$-$1000~Ryd range and $c$ is the speed of light, and 
   \begin{equation}
      {\dot P_{scatt}} = \frac {L_{ion}}{c}(1 - e^{-\tau_{T}})  \,,    
   \end{equation}
using the ionising luminosity $L_{ion}$ this time, and $\tau_{T}$ (the optical depth for Thomson scattering) is given by
   \begin{equation}
      \tau_{T} = \sigma_{T} {N_{H}} \,    
   \end{equation}
where $\sigma_{T}$ is the Thomson cross-section.
This gives an expression for the volume filling factor:
   \begin{equation}
   {C_v} \sim \frac {{({\dot P_{abs}} + {\dot P_{scatt}})} {\xi}}{1.23 m_p c L_{ion} v^2 \Omega} \,.  
   \end{equation}

The effect of the absorber is modelled using the \emph{xabs} ionised absorption model in SPEX 2.00, which includes line and edge absorption self-consistently and allows the column density, ionisation parameter $\xi$, outflow velocity, turbulent velocity and elemental abundances of an ionised absorber to be specified. For each absorbing phase, N$_{\rm H}$, log $\xi$ and v$_{\rm out}$ are taken from Table~\ref{phase_properties}, the turbulent velocity is set to 100~km~s$^{\rm -1}$ and the elemental abundances are assumed to be Solar \citep{anders1989}.

\begin{figure}
   \centering
   \includegraphics[width=8.5cm]{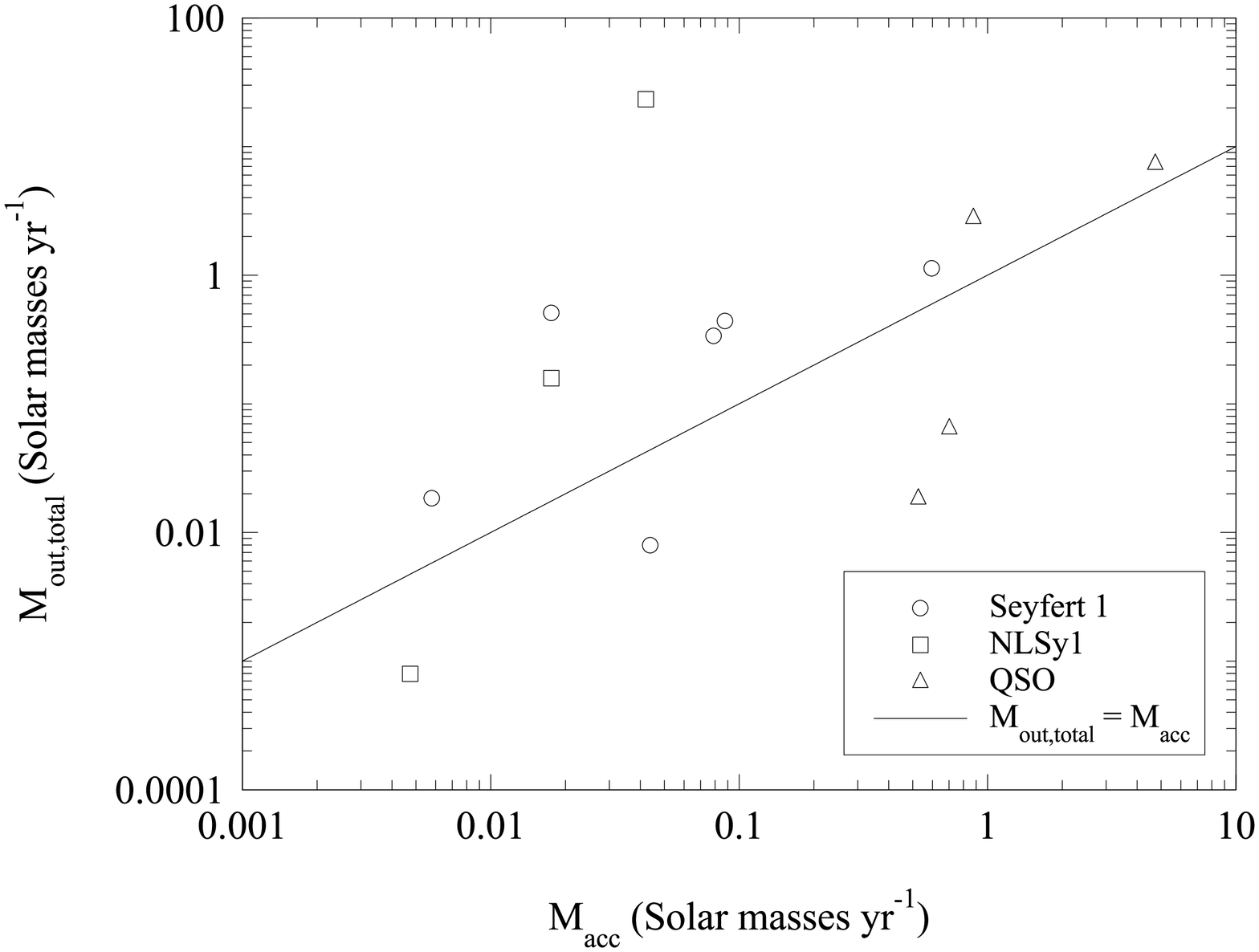}
      \caption{Total mass outflow rate of each AGN warm absorber ${\dot M}_{out,total}$ plotted against ${\dot M}_{acc}$, the mass accretion rate of the AGN in each case. The line where ${\dot M}_{out,total}$=${\dot M}_{acc}$ is plotted for comparison.}
         \label{mout_macc}
   \end{figure}

The estimated values for $C_v$ of each phase are listed in Table~\ref{phase_wa_energetics}; the combined volume filling factors for all measured phases in each object are never more than $\sim$ 8\%. Using these values, we then obtained the mass outflow rates (${\dot M}_{out}$) for each individual phase which are also given in Table~\ref{phase_wa_energetics}. For comparison, we include values for the mass accretion rate of each AGN calculated according to
   \begin{equation}
      {\dot M}_{acc} = \frac {L_{bol}}{c^2 \eta}  \,,    
   \end{equation}
where $\eta$ is the accretion efficiency (we assume $\eta$ = 0.1 for each object), and also the ratio ${\dot M}_{out,total}$/${\dot M}_{acc}$ where ${\dot M}_{out,total}$ is the total mass outflow rate summed over all phases. ${\dot M}_{out,total}$ is plotted against ${\dot M}_{acc}$ for each galaxy (with outflowing warm absorber components) in Fig.~\ref{mout_macc}. The outflow rate is greater than the accretion rate in 9 out of 13 objects (2/3 NLSy1s, 5/6 Seyfert 1s, and 2/4 QSOs). We note that, in the case of the two QSOs whose warm absorbers have been claimed to be accretion disc winds (PG0844+349 and PG1211+143; \citealt{pounds20030844,pounds20031211}), the total mass outflow rates are about 4\% and 9\% of the mass accretion rates respectively.

Thomson scattering is only expected to make a significant contribution to the momentum transfer in outflow phases with high absorbing columns; in reality, because a given phase might absorb very little radiation, scattering can be responsible for a greater proportion of the total momentum exchange than one might assume. Even so, in all objects except the two PG quasars, taking Thomson scattering into account increases the overall mass outflow rate, summed over all phases, by no more than 8\%. It turns out that the phases in which scattering has the greatest effect are generally those with the lowest mass outflow rates; the percentage of the total momentum transfer which is due to Thomson scattering is listed for each phase in Table~\ref{phase_wa_energetics}. In the cases of PG0844+349 and PG1211+143, which have very high absorbing columns, most of the momentum transfer occurs through scattering, and taking it into account gives mass outflow rates respectively $\sim$ 20 and 4 times those calculated if it is assumed that momentum is only transferred through absorption.

Are these outflows significant in energetic terms? The kinetic energy carried via the outflow per unit time, i.e. the kinetic luminosity, can be estimated as
   \begin{equation}
      L_{KE} = \frac{1}{2} {\dot M}_{out} v^2  
   \end{equation}
where we assume that the observed velocity $v$ is representative of the main mass of the outflow. We list values for L$_{\rm KE}$ for each phase in Table~\ref{phase_wa_energetics} alongside the percentage of L$_{\rm bol}$ that L$_{\rm KE}$ represents. In all cases, the kinetic luminosity of the warm absorber phases represents less than 1\% of the bolometric luminosity of the AGN, and so the ionised outflows cannot be playing any major r\^{o}le in the energetics of the system.

   \begin{table*}
    
      \caption[]{Mass outflow rates calculated for each \emph{outflowing} warm absorber phase using the parameters quoted in Table~\ref{phase_properties} (Markarian 766 does not have an outflowing warm absorber, but the ionising luminosity is listed as we use it elsewhere): object name, log ionisation parameter (log $\xi$; erg cm s$^{\rm -1}$), log 1$-$1000~Ryd ionising luminosity (log L$_{\rm ion}$; erg s$^{\rm -1}$), X-ray to optical spectral index ($\alpha_{ox}$), percentage of total momentum transfer due to Thomson scattering (\% ${\rm \dot M}_{\rm scat}$), percentage volume filling factor (\% $C_v$), mass outflow rate per phase (${\rm \dot M}_{\rm out}$; $M_\odot$ yr$^{\rm -1}$), mass accretion rate (${\rm \dot M}_{\rm acc}$; $M_\odot$ yr$^{\rm -1}$), ratio of total mass outflow rate (summed over all phases) to mass accretion rate (${\rm \dot M}_{\rm out,total}$/${\rm \dot M}_{\rm acc}$), kinetic luminosity of outflow (log L$_{\rm KE}$; erg s$^{\rm -1}$), percentage of L$_{\rm bol}$ represented by L$_{\rm KE}$ (\% of L$_{\rm bol}$).}
         \label{phase_wa_energetics}  
   \centering
         \begin{tabular}{llllllllllr}
            \hline
            \noalign{\smallskip}
Object & log $\xi$ & log L$_{\rm ion}$ & $\alpha_{ox}$ & \% ${\rm \dot M}_{\rm scat}$ & \% $C_v$ & ${\rm \dot M}_{\rm out}$ & ${\rm \dot M}_{\rm acc}$ & ${\rm \dot M}_{\rm out,total}$/${\rm \dot M}_{\rm acc}$ & log L$_{\rm KE}$ & \% of L$_{\rm bol}$  \\
            \noalign{\smallskip}
            \hline
            \noalign{\smallskip}
MR2251-178	& 2.9	& 45.5	& 1.24  & 35   & 7.9	  & 0.37    & 0.88   & 3.2    & 39.9 & 0.0001 \\
        	& 0.68	& 	&       & 0.33 & 0.31	  & 2.4	    &        &        & 40.7 & 0.001 \\
PG0844+349	& 3.7	& 45.0	& 1.45  & 95   & 0.031   & 0.018   & 0.53   & 0.035  & 43.4 & 0.8 \\
PG1211+143	& 3.4	& 44.9	& 1.53  & 89   & 0.14	  & 0.051   & 0.70   & 0.092  & 43.0 & 0.2 \\
        	& 1.7	& 	&       & 5.1  & 0.00072 & 0.013   &        &        & 42.4 & 0.06 \\
IRAS 13349+2438	& 0	& 45.0	& 1.31  & 0.22 & 0.30	  & 7.4     & 4.7    & 1.6    & 41.6 & 0.002 \\
NGC 4593	& 2.61	& 43.7	& 1.15  & 22   & 1.2	  & 0.0032  & 0.0058 & 3.2    & 38.2 & 0.0005 \\
        	& 0.5	& 	&       & 0.19 & 0.049	  & 0.015   &        &        & 38.8 & 0.002 \\
NGC 3783	& 1.1	& 44.1	& 1.28  & 1.9  & 0.63	  & 0.28    & 0.079  & 4.3    & 40.7 & 0.01 \\
        	& 2.3	& 	&       & 19   & 1.3	  & 0.035   &        &        & 39.8 & 0.001 \\
        	& 2.9	& 	&       & 50   & 3.8	  & 0.026   &        &        & 39.7 & 0.001 \\
Markarian 509	& 1.76	& 45.1	& 1.38  & 4.1  & 4.7	  & 1.1     & 0.60   & 1.9    & 40.2 & 0.0004 \\
NGC 7469	& 2.1	& 44.4	& 1.25  & 6.1  & 0.086	  & 0.0080  & 0.044  & 0.18   & 39.2 & 0.0006 \\
NGC 3516	& 0.78	& 43.6	& 1.44  & 1.1  & 7.2	  & 0.50    & 0.018  & 29     & 39.8 & 0.006 \\
NGC 5548	& 2.69	& 44.5	& 1.24  & 30   & 5.5	  & 0.054   & 0.088  & 5.0    & 39.2 & 0.0003 \\
        	& 1.98	& 	&       & 7.0  & 1.6	  & 0.11    & 	     &        & 39.8 & 0.001 \\
        	& 0.4	& 	&       & 0.18 & 0.15	  & 0.27    & 	     &        & 39.9 & 0.001 \\
MCG-6-30-15	& 1.25	& 43.7	& 0.76  & 1.5  & 6.9	  & 0.16    & 0.018  & 9.0    & 39.1 & 0.001 \\
        	& 2.5	& 	&       & 19   & 0.06	  & 0.0010  & 	     &        & 39.1 & 0.001 \\		
Markarian 766   & 0.7	& 44.1	& 1.02  & 0.51 &          &         &        &        &      &        \\
NGC 4051	& 1.4	& 42.3	& 1.20  & 2.4  & 0.44	  & 0.00080 & 0.0047 & 0.17   & 37.6 & 0.0002 \\		
Ark 564	        & 0	& 45.2	& 1.06  & 0.16 & 1.9	  & 23	    & 0.042  & 550    & 41.2 & 0.07  \\
                & 2	& 	&       & 5.1  & 5.9	  & 0.71    & 	     &        & 39.7 & 0.002 \\
            \noalign{\smallskip}
            \hline
         \end{tabular}        
  \end{table*}

\subsection{The origins of warm absorbers}

The outflow speed can give us a useful clue to the origins of warm absorbers. If we assume that the outflow has to obtain a speed greater than or equal to the escape velocity v$_{\rm esc}$, where
   \begin{equation}
      v_{esc} = \sqrt{\frac{2 G M}{R}}  
   \end{equation}
and $G$ is the gravitational constant, $M$ is the mass of the black hole and $R$ is the distance of the bulk of the gas from the black hole, then we can estimate a minimum distance of the warm absorber from the central engine using the measured outflow speed (v$_{\rm out}$):
   \begin{equation}
      R \geq \frac{2 G M}{v_{out}^{2}}  \,. 
   \end{equation}

The maximum distance of the base of the warm absorber can also be estimated. Assuming that most of the mass of the absorber with ionisation parameter $\xi$ is concentrated in a relatively thin layer of depth $\Delta$$r$ (since the density falls off rapidly with radius), $\Delta$$r$ has to be less than or equal to the distance R from the central engine:
   \begin{equation}
\label{r_delta_r}
      \frac{{\Delta}r}{R} \leq 1 \,.
   \end{equation}

The observed line-of-sight absorbing column $N_H$ is a function of the density $n(R)$ of the material at ionisation parameter $\xi$, its volume filling factor $C_v$ and $\Delta$$r$;
   \begin{equation}
\label{col}
      N_{H} \sim n(R) C_v {\Delta}r  \,,
   \end{equation}
which can be combined with the expression for the ionisation parameter of the shell
   \begin{equation}
\label{xi}
      \xi = \frac{L_{ion}}{n(R) R^2}
   \end{equation}
to give
   \begin{equation}
      \frac{{\Delta}r}{R} \sim \frac{\xi R N_H}{L_{ion} C_v(R)}  \,.
   \end{equation}

Applying the condition that $\Delta$$r$/$R$ $\leq$ 1, one then gets
   \begin{equation}
      R \leq \frac{L_{ion} C_v(R)}{\xi N_H}  \,.
   \end{equation}

\begin{figure*}
   \centering
   \includegraphics[width=15cm]{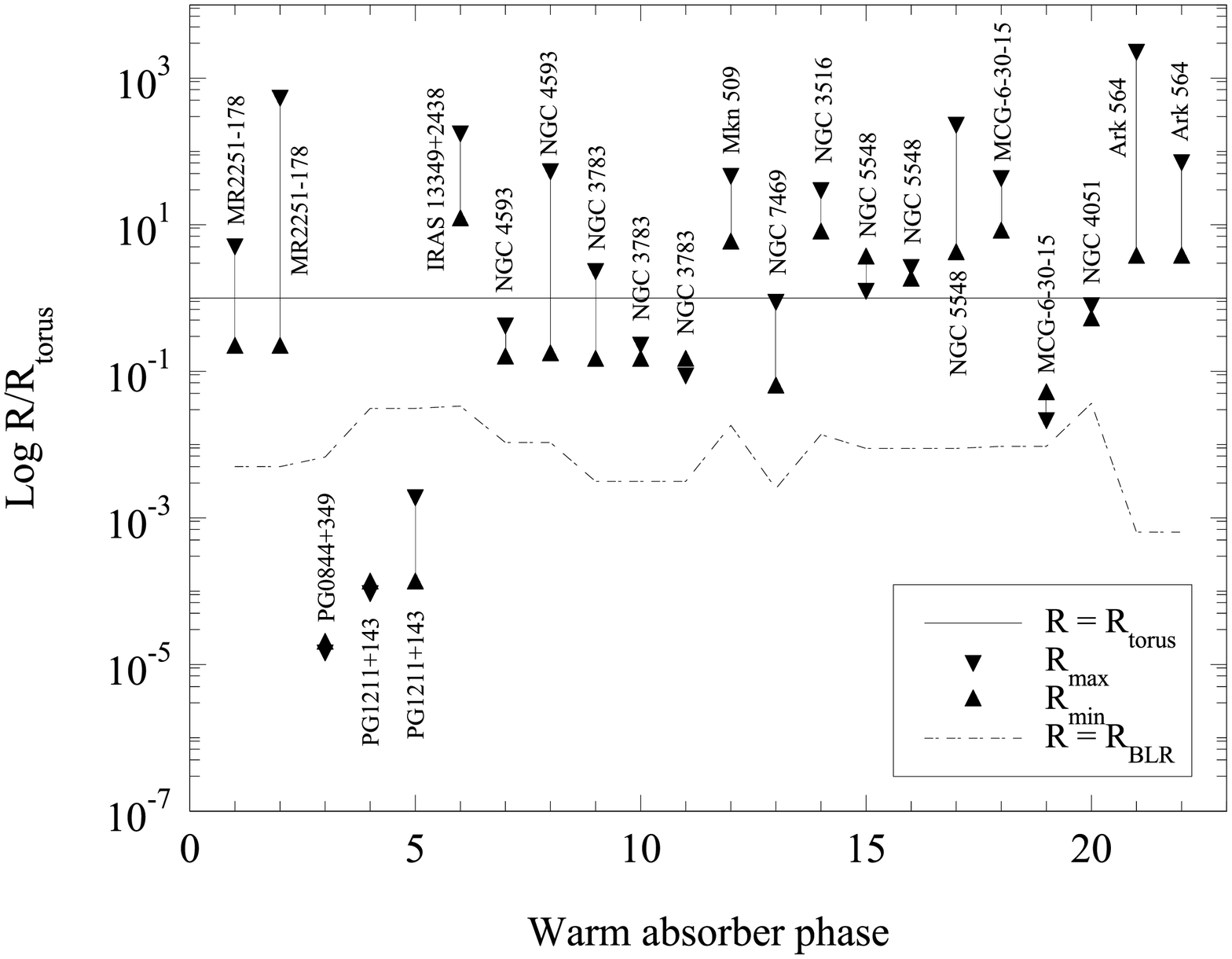}
      \caption{The estimated minimum (R$_{\rm min}$) and maximum (R$_{\rm max}$) distances of the warm absorber phases from the central engine as a ratio to the torus distance R$_{\rm torus}$, plotted for each outflowing warm absorber phase in Table~\ref{distance_comp}, in the order in which they appear in the table. The ratio of the BLR distance (R$_{\rm BLR}$) to the torus distance is also plotted.}
         \label{wa_dists}
   \end{figure*}

The resulting minimum and maximum distances for each phase are listed in Table~\ref{distance_comp} alongside the distances of the BLR and torus for each object; since both r$_{\rm min}$ and r$_{\rm max}$ (through $C_v$) are dependent upon $v$, it is only possible to calculate them for outflowing phases. The BLR distances were taken from the literature, except in the cases of IRAS 13349+2438, MCG -6-30-15 and Markarian 766 where they were estimated using r$_{\rm BLR} \propto$ L$_{\rm Bol}^{\rm 0.5}$ \citep[e.g.][]{wandel1999}, with the BLR distance of NGC 5548 (a reverberation measurement from \citealt{wandel1999}) used to normalise the proportionality. Torus distances were estimated using the approximate relation given in \citet{krolik2001} that the inner edge of the torus is $\sim$ 1~$\times$~L$_{\rm ion,44}^{\rm 0.5}$ pc, where L$_{\rm ion,44}$ is the 1$-$1000~Ryd luminosity L$_{\rm ion}$ in units of 10$^{\rm 44}$ erg s$^{\rm -1}$. The ratios of r$_{\rm max}$, r$_{\rm min}$ and r$_{\rm BLR}$ to r$_{\rm torus}$ are plotted in Fig.~\ref{wa_dists} for each of the phases in Table~\ref{distance_comp}.

   \begin{table*}
      \caption[]{Comparison of distances within the AGN environment: object name, log ionisation parameter of each phase (log $\xi$; erg cm s$^{\rm -1}$), black hole mass (M$_{\rm BH}$; 10$^{\rm 7}$ $\rm M_\odot$), distance of BLR from central engine (r$_{\rm BLR}$; pc), distance of torus from central engine (r$_{\rm torus}$; pc), minimum distance of warm absorber from central engine (r$_{\rm min}$; pc), maximum distance of warm absorber from central engine (r$_{\rm max}$; pc), the ratios of the minimum and maximum distances to the BLR and torus distances respectively (r$_{\rm min}$/r$_{\rm BLR}$, r$_{\rm max}$/r$_{\rm BLR}$, r$_{\rm min}$/r$_{\rm torus}$, r$_{\rm max}$/r$_{\rm torus}$).}
         \label{distance_comp}  
   \centering
         \begin{tabular}{llllllllllr}
            \hline
            \noalign{\smallskip}
Object & log $\xi$ & M$_{\rm BH}$ & r$_{\rm BLR}$ & r$_{\rm torus}$ & r$_{\rm min}$ & r$_{\rm max}$ & r$_{\rm min}$/r$_{\rm BLR}$ & r$_{\rm max}$/r$_{\rm BLR}$ & r$_{\rm min}$/r$_{\rm torus}$ & r$_{\rm max}$/r$_{\rm torus}$ \\
            \noalign{\smallskip}
            \hline
            \noalign{\smallskip}
MR2251-178	& 2.9   & 0.83$^{\mathrm{a}}$ & 0.027$^{\mathrm{b}}$   & 5.4     & 1.1	   & 29     & 43     & 1100	 & 0.21 & 5.3 \\
	        & 0.68	&		      &                        &         & 1.1     & 3000   & 43     & 110000  & 0.21 & 570 \\
PG0844+349	& 3.7   & 2.7$^{\mathrm{c}}$  & 0.020$^{\mathrm{c}}$   & 3.0 & 5.9 $\times$ 10$^{\rm -5}$ & 4.6 $\times$ 10$^{\rm -5}$ & 0.0029 & 0.0023 & 2.0 $\times$ 10$^{\rm -5}$ & 1.5 $\times$ 10$^{\rm -5}$ \\
PG1211+143	& 3.4   & 2.36$^{\mathrm{c}}$ & 0.085$^{\mathrm{c}}$   & 2.7 & 0.00035 & 2.7$\times$ 10$^{\rm -4}$ & 0.0042 & 0.0032 & 0.00013  & 9.8 $\times$ 10$^{\rm -5}$ \\
	        & 1.7	&		      &                        &        & 0.00035 & 0.0054 & 0.0042 & 0.064 & 0.00013 & 0.0020 \\
                & -0.9  & 2.36                & 0.085                  & 2.7     & $-$     & $-$    & $-$   & $-$     & $-$   & $-$ \\
IRAS 13349+2438	& 2.25  & 80$^{\mathrm{d}}$   & 0.11$^{\mathrm{e}}$    & 3.3     & $-$	   & $-$    & $-$   & $-$     & $-$   & $-$ \\
	        & 0	&		      &                        &         & 39      & 620    & 350   & 5500    & 12    & 180 \\
NGC 4593	& 2.61  & 0.2$^{\mathrm{f}}$  & 0.0076$^{\mathrm{f}}$  & 0.71    & 0.11    & 0.31   & 14    & 41      & 0.15  & 0.44 \\
	        & 0.5	&		      &                        &         & 0.12    & 40	    & 16    & 5300    & 0.17  & 56 \\
NGC 3783	& 1.1   & 1.1$^{\mathrm{g}}$  & 0.0038$^{\mathrm{g}}$  & 1.2	 & 0.17    & 2.9    & 45    & 770     & 0.14  & 2.4 \\
	        & 2.3	&		      &                        &         & 0.17    & 0.29   & 45    & 77      & 0.14  & 0.24 \\
	        & 2.9	&		      &                        &         & 0.17	   & 0.11   & 45    & 29      & 0.14  & 0.092 \\
Markarian 509   & 1.76  & 9.5$^{\mathrm{g}}$  & 0.067$^{\mathrm{g}}$   & 3.6	 & 20	   & 180    & 310   & 2700    & 5.6   & 49 \\
NGC 7469	& 2.1   & 0.76$^{\mathrm{g}}$ & 0.0042$^{\mathrm{g}}$  & 1.67	 & 0.10    & 1.6    & 24    & 370     & 0.061 & 0.93 \\
NGC 3516	& 0.78  & 2.3$^{\mathrm{h}}$  & 0.0088$^{\mathrm{i}}$  & 0.64	 & 4.9     & 20	    & 560   & 2300    & 7.7   & 31 \\
NGC 5548	& 2.69  & 6.8$^{\mathrm{g}}$  & 0.015$^{\mathrm{g}}$   & 1.7	 & 6.0     & 2.3    & 400   & 150     & 3.5   & 1.3 \\
        	& 1.98	&		      &                        &         & 3.0	   & 4.8    & 200   & 310     & 1.8   & 2.8 \\
        	& 0.4	&		      &                        &         & 7.0	   & 420    & 460   & 27000   & 4.0   & 240 \\
MCG-6-30-15	& 1.25  & 1.5$^{\mathrm{j}}$  & 0.0068$^{\mathrm{e}}$  & 0.72    & 5.7     & 33	    & 840   & 4800    & 7.9   & 46 \\
	        & 2.5	&		      &                        &         & 0.036   & 0.016  & 5.2   & 2.4     & 0.050 & 0.023 \\
Markarian 766   & 0.7   & 1.9$^{\mathrm{k}}$  & 1.1$^{\mathrm{e}}$     & $-$	 & $-$     & $-$    & $-$   & $-$     & $-$   & $-$ \\
NGC 4051	& 1.4   & 0.14$^{\mathrm{g}}$ & 0.0054$^{\mathrm{g}}$  & 0.15	 & 0.075   & 0.12   & 14    & 23      & 0.50  & 0.84 \\
Ark 564	        & 0 & 3.7$^{\mathrm{l}}$ & $\leq$0.0025$^{\mathrm{m}}$ & 3.9	 & 14	   & 9400   & 5600  & 3700000 & 3.6   & 2400 \\
	        & 2	&		      &                        &         & 14      & 290    & 5600  & 120000  & 3.6   & 75 \\
            \noalign{\smallskip}
            \hline
         \end{tabular}        
\begin{list}{}{}
\item[$^{\mathrm{a}}$] \citealt{morales2002}; $^{\mathrm{b}}$ \citealt{monier2001}; $^{\mathrm{c}}$ \citealt{kaspi2000rblr}; $^{\mathrm{d}}$ \citealt{brandt1997}; $^{\mathrm{e}}$ calculated from L$_{\rm Bol}$ as described in text; $^{\mathrm{f}}$ \citealt{santos1995}; $^{\mathrm{g}}$ \citealt{wandel1999}; $^{\mathrm{h}}$ \citealt{gebhardt2000}; $^{\mathrm{i}}$ \citealt{wanders1993}; $^{\mathrm{j}}$ \citealt{reynolds2000}; $^{\mathrm{k}}$ \citealt{mathur2001}; $^{\mathrm{l}}$ \citealt{matsumoto2004}; $^{\mathrm{m}}$ \citealt{shemmer2001}.
\end{list}
  \end{table*}

The results of this comparison are that the minimum and maximum warm absorber distances for the Seyferts and NLSy1s are all further out than the BLR, and mostly cluster around the distance of the torus. Placing the base of the warm absorber outflow as far out as the torus immediately rules out an origin in an accretion disc wind for these absorbers, and lends support to the torus wind concept \citep{krolik2001}. 

It is also worth investigating the effect of using the turbulent velocity ($v_{turb}$) of the warm absorber phases in the expression for the minimum warm absorber distance. Measured or fitted values for the turbulent velocity (as opposed to guessed ones or upper limits, if a number is quoted at all) are only given for IRAS 13349+2438, NGC 3783, Markarian 509, MCG-6-30-15 and NGC 4051. These values, alongside the results of using them to calculate the minimum warm absorber distance (r$_{\rm min,vturb}$), are listed in Table~\ref{vturb_rmin} alongside their ratios to the BLR and torus distances and to the minimum radius calculated using the line of sight velocity. Using the turbulent velocity width to calculate the minimum radius gives greater distances in all cases except for the outflowing warm absorber phase in IRAS 13349+2438 and the low velocity phase of MCG-6-30-15. In each case, r$_{\rm min,vturb}$ lies far outside the BLR, and the values cluster around the distance of the torus. For this small set of turbulent velocities, then, we find that whether one calculates the minimum distance of the warm absorber using the outflow velocity or the turbulent velocity, the results are most consistent with the idea that the outflow originates in a torus wind.

   \begin{table*}
      \caption[]{The results of using the turbulent velocity rather than the outflow velocity to calculate the minimum distance of a warm absorber from the central engine: object name, turbulent velocity (v$_{\rm min,vturb}$; km~s$^{\rm -1}$ FWHM), minimum distance of warm absorber from central engine derived from turbulent velocity (r$_{\rm min,vturb}$; pc), the ratios of this minimum distance to the BLR and torus distances respectively (r$_{\rm min,vturb}$/r$_{\rm BLR}$, r$_{\rm min,vturb}$/r$_{\rm torus}$), and the ratio r$_{\rm min,vturb}$/r$_{\rm min}$.}
         \label{vturb_rmin}  
   \centering
         \begin{tabular}{lllllr}
            \hline
            \noalign{\smallskip}
Object & v$_{\rm turb}$ & r$_{\rm min,vturb}$ & r$_{\rm min,vturb}$/r$_{\rm BLR}$ & r$_{\rm min,vturb}$/r$_{\rm torus}$ & r$_{\rm min,vturb}$/r$_{\rm min}$ \\
            \noalign{\smallskip}
            \hline
            \noalign{\smallskip}
IRAS 13349+2438	           & 1430 & 3.4   & 30  & 1.0  & 0.086 \\
NGC 3783	           & 420  & 0.55  & 150 & 0.46 & 3.2   \\
Markarian 509              & 170  & 29    & 440 & 8.1  & 1.4   \\
MCG-6-30-15$^{\mathrm{a}}$ & 220  & 2.8   & 400 & 3.8  & 0.48  \\
	                   & 770  & 0.22  & 32  & 0.30 & 6.2   \\
NGC 4051	           & 350  & 0.097 & 18  & 0.65 & 1.3   \\
            \noalign{\smallskip}
            \hline
         \end{tabular}        
\begin{list}{}{}
\item[$^{\mathrm{a}}$] The two absorption phases in MCG-6-30-15 have different turbulent velocities and different outflow velocities.
\end{list}
  \end{table*}

But what about the \citet{kriss2003} and \citet{blustin2003} result that at least one phase of the warm absorber in NGC 7469 is actually within the BLR? If this suggestion is correct, it could imply that we are seeing the warm absorber in this source (and perhaps others) before it has been accelerated to the full escape speed, and therefore the gas can be closer to the central engine than we predict. This may be a reasonable suggestion, since we are seeing the bulk of the gas very close to the base of the outflow, and it will obviously take a finite distance to accelerate the outflow to escape velocity.

The situation is different with the quasars. The two objects with `normal' Seyfert-type outflows, MR2251-178 and IRAS 13349+2438, fit into the same pattern as the Seyferts. The two PG quasars, on the other hand, have minimum and maximum radii much closer to the central engine due to their extremely high outflow speeds, high columns and high average ionisation: the warm absorber of PG0844+349 is placed about 1.3$-$1.7 light-hours from the continuum source, and the corresponding limits for PG1211+143 are around 10 light-hours to 6 light-days. This makes an origin in an accretion disc wind most likely. 

\begin{figure}
   \centering
   \includegraphics[width=8.5cm]{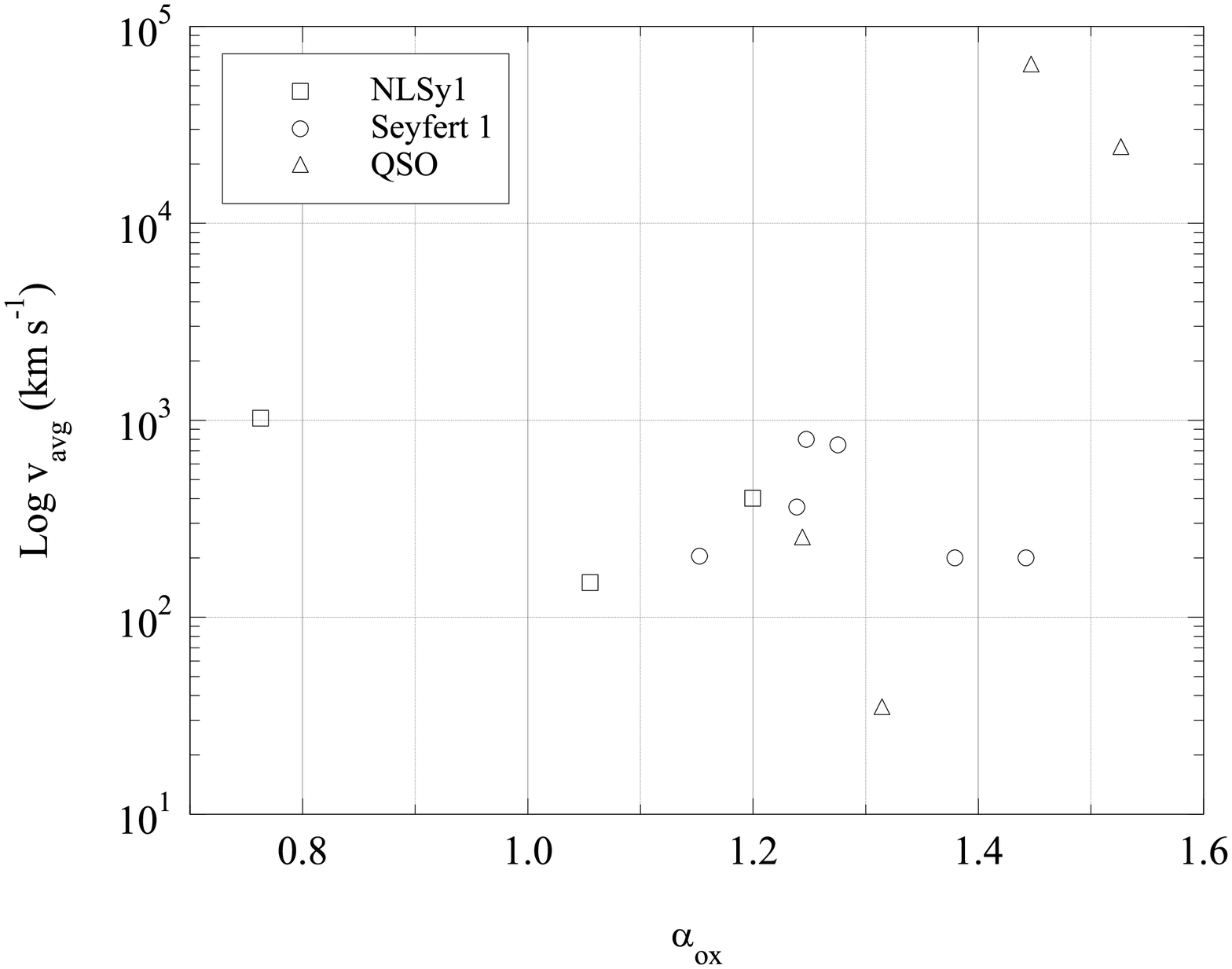}
      \caption{Log $v_{avg}$ (Table~\ref{average_properties}) versus $\alpha_{ox}$ (Table~\ref{phase_wa_energetics}) for the AGN with outflowing warm absorbers.}
         \label{vavg_aox}
   \end{figure}

In an accretion disc, the ability to launch a radiatively-driven wind is dependent upon the spectral shape; if the ratio of UV to X-ray flux is too low, the material in the outer layer of the disc is too highly ionised to be accelerated \citep[e.g.][]{proga2003}. Do the quasars with apparent accretion disc winds have higher $\alpha_{ox}$ indices than the other AGN? We estimated $\alpha_{ox}$ for each AGN in our sample having an outflowing warm absorber, using the fluxes at 2~keV (from our SEDS) and at 2500~\AA\, (from archival IUE spectra). The 2500~\AA\, fluxes were corrected for extinction due to the ISM of our Galaxy using the coefficients of \cite{savage1979}; the $\alpha_{ox}$ values are listed in Table~\ref{phase_wa_energetics}. Log $v_{avg}$ is plotted against $\alpha_{ox}$ in Fig.~\ref{vavg_aox}. The two PG quasars do indeed have the highest $\alpha_{ox}$ indices. NGC~3516, however, whose warm absorber is a torus wind, has a value of $\alpha_{ox}$ which is just as high.

There is evidence that these inner outflows also occur in nearby Seyferts. In NGC 3783, \citet{reeves2004} found evidence for a highly ionised phase of the warm absorber whose variability put it very close to the central engine. Crucially, the main spectral evidence for this phase were absorption lines in the 6$-$7~keV band, where resolution and signal-to-noise are much lower than in the soft X-rays. This very highly ionised gas is hard to observe in soft X-ray spectra, and this may be why it has not been reported in more objects. There could therefore be significant columns of such gas which have previously been unobserved.

So where does this take our understanding of the origins of warm absorbers? We seem to be dealing with two different basic types of warm absorber - the type we see in nearby Seyferts, which seems to be most consistent with a torus wind, and the rather more extreme variety observed in the two PG quasars and in the hard X-ray spectrum of NGC 3783, which could correspond to an accretion disc wind.

One interesting point about the extreme outflows is that their presence is not a simple function of luminosity, as the two most luminous quasars we consider here have low velocity, low column Seyfert-type outflows. It is not a simple function of spectral shape, either, since Seyferts can have an $\alpha_{ox}$ index as high as that of the accretion disc wind quasars. It is possible that the apparent lack of accretion disc winds in nearby Seyferts is an observational effect, and that they are only detectable in absorption above 6~keV; another explanation might be that accretion disc winds in Seyferts travel almost perpendicular to our line of sight (as discussed in the next Section). Alternatively, there could be a difference either in the availability of matter for the outflow, or in the process which can drive it. Within the unified model of AGN, this must mean either that there is some difference in the structure of the inner regions of the AGN environment, or that there is something different about the accretion process which can drive an accretion disc wind.

\subsection{The relationship between UV and X-ray absorbers}

It is now worth asking how this relates to the knowledge gained about AGN outflows from studies of UV absorbers and emitters. The higher resolution of current astronomical UV spectrometers allows us to gain quite a detailed view of the kinematics of UV absorbers. The first problem to solve, before making a comparison between outflows in the two wavebands, is the precise relationship between UV and X-ray absorption. Different conclusions have been reached about different sources: at least some of the components of the X-ray and UV absorbers have been claimed to have a common origin in NGC~3783 \citep{kaspi2002,gabel2003}, NGC~3516 \citep{kraemer20023516}, NGC~7469 \citep{kriss2003,blustin2003} and Ark~564 \citep{romano2002}. Conversely, the UV and X-ray absorption apparently originate in different phases of gas (albeit at similar velocities) in Markarian~509 \citep{kraemer2003}.

The emerging picture is that some but not all of the UV velocity phases can be seen in the X-ray absorber, and some but not all of the X-ray absorber phases can be seen in the UV; the deciding factors are the column density and ionisation parameters of the phases. Plasma which absorbs in the UV will contain ions that are observable in the soft X-ray band, although the line-of-sight column of the plasma may not be high enough for the X-ray transitions to be visible to current instrumentation. Plasma absorbing in the X-rays, so long as it is not too highly ionised, can contain ions with transitions in the UV. Working out exactly which of the UV and X-ray phases observed in any given object coincide ideally requires simultaneous high resolution spectroscopic observations in both bands.

UV absorbers generally have more than one discrete velocity component \citep{crenshaw1999}. It can be hard to work out whether the gas phases giving rise to these components actually have different ionisation parameters, due to the limited number of spectral lines visible in certain observable wavebands. Nevertheless, two basic interpretations are possible; either we are seeing different velocity streams with different origins, or we are seeing a single accelerating (or decelerating) outflow with a multiply-peaked density profile. This latter case could imply that the ionised absorber is not a single time-steady outflow, but is the result of a series of ejection events. If the UV wind originated from the torus, for example, then this could be due to major variations over time in the AGN luminosity (i.e. high and low states, or flares) giving rise to peaks and troughs in the amounts of material ablated from the torus. The multiple phases would thus contain an imprint of the accretion history of the AGN. 

\citet{kraemer20023516} note that there is a lack of evidence of radial acceleration in AGN absorbers. Their analysis of UV absorbers in NGC~3516 does find that the different kinematic phases are moving faster the further they are from the nucleus (the X-ray absorber appears to be associated with the slowest moving phases closest to the nucleus), but they remark that this is not necessarily the sign of an accelerating outflow as the phases could have originated at different distances from the nucleus. As we discussed earlier, faster moving gas in a radiatively accelerated outflow would be expected to be more highly ionised - so are the X-ray absorbers moving faster than the UV absorbers? The observational evidence tends to favour the opposite scenario, with (at least some) of the UV phases moving faster than the X-ray phases, for instance in NGC~3516 \citep{kraemer20023516} and NGC~7469 \citep{kriss2003}. This may, again, be due to the decreasing column problem; UV spectrometers are able to detect much lower columns of gas than current X-ray instruments.

There is evidence of increasing velocity with distance from the nucleus in the UV NLR, interpreted as the result of radiative acceleration, in the emission line dominated objects NGC~1068 and NGC~4151. The ionisation cones forming the NLRs can be modelled as outflowing bicones at least partially evacuated along their axes \citep{crenshaw20001068,crenshaw20004151}. By measuring the velocities of individual knots or clouds in the ionisation cones, higher speeds were observed at greater distances from the nucleus, and then a decrease in speed further out, interpreted as being due to interactions with a patchy ambient medium. In the case of NGC~4151 the acceleration of emission line knots continues up to 160~pc away from the nucleus, with decelerated clouds approaching the mean systemic velocity at 290~pc. In NGC~1068, UV continuum radiation has been observed from an apparent shock front where an emission line knot is being decelerated \citep{kraemer20001068}.

So, the gas in the UV NLR (which has been associated with the X-ray NLR, e.g. \citealt{kinkhabwala2002}) does appear to be an outflow, plausibly accelerated by radiation from the central engine. \cite{crenshaw20004151} interpret the apparent evacuation of the NLR ionisation cone along its axis as possibly being due to the radio jet in this object; interestingly, a hollow bicone is the geometry one would expect from a torus wind. The observed column and ionisation level of the warm absorber would therefore be a function of viewing angle into the central engine, and there would be certain sightlines (along the axis of the cone) where no warm absorber was visible at all.

Can the `inner' outflow, which could be an accretion disc wind or an outflowing BLR, be seen in UV? \citet{hutchings2001} found that the P-Cygni profile of \ion{O}{vi} observed whilst NGC~3516 was in a low state was evidence that the BLR was actually an outflowing wind. An obvious source for the material in an outflowing BLR is the accretion disc wind. This may provide an explanation of why there is no trace of the accretion disc wind in the X-ray warm absorbers of nearby Seyferts. It is possible that the BLR is outflowing at a large angle to the axis of the system (the turbulent velocity of BLR gas is far higher than its outflow velocity along our line of sight) making it impossible to observe in absorption. This geometry is consistent with the radiatively driven accretion disc wind models of \citet{proga2003}, where material is driven off the disc at a very low angle to the disc surface. In this scenario, the observational evidence that accretion disc winds are visible along our line of sight in certain quasars raises the interesting possibility that either the wind is not radiatively driven in such objects, or that the geometry of the central engine is somewhat different to that in nearby Seyferts.

The general picture that we obtain from the X-ray warm absorbers in Seyfert-type objects - namely that there appear to be two types of AGN outflow with different origins - is reflected in the findings of a recent survey of intrinsic UV absorption in 56 PG quasars. \citet{laor2002} compared UV absorption in Soft X-ray Weak Quasars (SXWQs) and non-SXWQs. SXWQ UV outflows have a very high column (the more luminous SXWQs are Broad Absorption Line (BAL) Quasars) and presumably cause the soft X-ray spectra to be highly absorbed. Non-SXWQs, on the other hand, have UV absorbers which resemble those in nearby Seyferts. They concluded that the non-SXWQ outflows are launched at ten times the distance from which SXWQ outflows originate.

\subsection{The influence of warm absorbers on the AGN host galaxy}

The mass carried in the warm absorber must be ending up somewhere. Does it have an important effect on the host galaxy? One firstly needs to know if the ionised outflow is a short-lived phenomenon, or whether it exists for long enough to have a significant influence on the mass budget of the AGN. 

In the absence of replenishment of warm absorber material, how long would the observed columns of gas last? If one approximates the warm absorber as a slab of gas of thickness $\Delta$$r$ lying at a distance $r$ from the absorber, with a mass density $\rho$ and subtending a solid angle $\Omega$, the mass $M_{WA}$ of the absorber is given by:
   \begin{equation}
      M_{WA} = r^2 {\rho} {\Delta}{r} \Omega  \,. 
   \end{equation}
Since the mass density $\rho$ $\sim$ 1.23$m_p$${\bar n}$, where ${\bar n}$ is the average ion number density of the column and $m_p$ is the proton mass (as before, the factor of 1.23 corrects for a $\sim$ 25\% He, 75\% H by mass elemental composition),
   \begin{equation}
      M_{WA} \sim 1.23 m_p {\bar n} r^2 {\Delta}{r} \Omega  \,, 
   \end{equation}
and using the fact that the equivalent hydrogen column N$_{\rm H}$ is given by
   \begin{equation}
      N_{H} = {\bar n} {\Delta}{r} \,,
   \end{equation}
the mass of the absorber can be estimated as
   \begin{equation}
      M_{WA} \sim 1.23 m_p N_H r^2 \Omega  \,.
   \end{equation}

Assuming a solid angle of 1.6, as before, using the average N$_{\rm H}$ for the Seyfert 1 galaxies in our sample (10$^{\rm 21.7}$ cm$^{\rm -2}$), and assuming that the warm absorber slab is at the distance of the torus ($\sim$ 1~pc from the central engine), we estimate a representative Seyfert warm absorber mass of approximately 80 $M_\odot$. If the mass outflow rate was of the order of a solar mass per year, the warm absorber gas would not last out a century; even if the mass outflow rates are two or three orders of magnitude smaller than this, due to a low volume filling factor, the warm absorber would still be extremely short-lived compared to the AGN itself. This would have implications for the observability of ionised outflows; if they only lasted 10$^{\rm 2}$$-$10$^{\rm 5}$ years, there would be a low probability of observing them during the total AGN lifetime of perhaps 10$^{\rm 8}$ years. This means that the warm absorber column must be fed constantly by some long-lived source of matter - presumably, in the case of nearby Seyferts, the torus.

The median accretion rate for the Seyferts and NLSy1s in our sample is $\sim$ 0.04 M$_\odot$ yr$^{\rm -1}$, while the median outflow rate is $\sim$ 0.3 M$_\odot$ yr$^{\rm -1}$. Assuming that the ratio of mass outflow through the warm absorber to accretion through the disc is approximately constant throughout the lifetime of an AGN and that a typical Seyfert builds up a 10$^{\rm 7}$ M$_\odot$ black hole, it will push around 8 $\times$ 10$^{\rm 7}$ M$_\odot$ through its ionisation cones over the course of its existence.

Having established that the outflow is likely to be supplied with material continously throughout the life and evolution of an AGN, and
that it could potentially be returning $\sim$ 10$^{\rm 8}$ M$_\odot$ of material into the surrounding galaxy, we have to ask what effect it might have. Firstly, is this a significant mass of material compared to the central regions of the galaxy? Since the outflow is expected to expel $\sim$ 8 times more material than the black hole accretes, and that the black hole itself is typically only 0.15\% of the mass of its host galaxy spheroid \citep{merritnferrarese01}, the total mass outflowing over the history of the AGN is around 1\% of the mass of the spheroid. The hot ISM of a luminous elliptical galaxy typically represents of order a tenth of the mass of the stellar component \citep{osullivanetal03}. However, in the bulges of early type spiral galaxies the hot gas is a much smaller component, representing e.g. $\sim$ 10$^{-4}$ of the bulge mass in M~81 \citep{pageetal03}. The AGN wind is therefore likely to contribute a significant component of the hot ISM in galaxy spheroids, and could at times be the dominant provider of hot gas in bulges which host 10$^{\rm 7}$ M$_\odot$ black holes. The outflow speeds of Seyfert warm absorbers, when thermalised, would give rise to gas at the appropriate temperature.

A very pertinent question here is whether the gas stays in the galaxy itself, or escapes into the IGM. In part this depends simply on the mass of the host galaxy: more massive host galaxies will have higher escape velocities and therefore will be able to retain the gas from higher velocity outflows. However it will also depend on how much material the outflowing gas encounters along its trajectory. The \citet{crenshaw20004151} study of the UV NLR outflow in NGC~4151 showed that the outflow had decelerated to a standstill by the time it had reached about 290~pc away from the central engine; this distance is of the order of half the vertical thickness of a nearby spiral galaxy. This implies that the outflow material, in this case, is reaching the `top' of the galaxy but may not be moving out into the IGM.

\citet{kauffmann2003} find, using data from the Sloan Digital Sky Survey (SDSS), that AGN tend to show evidence for recent bursts
of star formation. For ionised outflows to contribute to this, they would have to contain or encounter regions of sufficiently high
density, perhaps through shocks of the type suggested to be present in NGC~1068 \citep{crenshaw20001068}. This is certainly an area which
deserves more attention. Considering, also, that the amount of mass removed from the active nucleus is of the same order as the mass accreted by the black hole itself, it is clear that there is a potential competition for the mass reservoir; the removal of mass through the ionised outflow may be shortening the lifetime of the AGN. 

\section{Conclusions}

In this paper, we collate results of high resolution X-ray spectroscopy of a sample of Seyfert 1 type AGN, and survey the phenomenological and physical properties of their warm absorbers. This is to some extent a risky exercise, since the spectra were analysed by many different authors, whose analysis methods were by no means uniform. Nevertheless, a coherent picture of the warm absorber phenomenon does emerge from the dataset as a whole.

We show that, in nearby Seyfert galaxies, the kinetic luminosities of the warm absorber outflows represent less than 1\% of the bolometric luminosities of the AGN. We also demonstrate that these warm absorbers most probably originate in an outflow from the dusty torus. It is, however, possible that highly ionised absorption seen above 6~keV could originate much closer to the nucleus, and this may be the best place to look if one is interested in searching for accretion disc winds in Seyfert galaxies. 

Our analysis shows the importance of taking into account the volume filling factors of the different ionisation phases in warm absorbers, and the relatively low volume filling factors we derive (less than 10\%) imply that some other gaseous medium (which does not absorb in the soft X-ray band) takes up the rest of the volume. This unseen medium could contain UV absorbing phases as well as very highly ionised, low density gas which - if its overall column was high enough - could produce absorption features above 6~keV.

We find that the warm absorbers of nearby quasars can be consistent with the torus wind type outflow seen in Seyferts. Indeed, comparison of similar warm absorbers in Seyfert galaxies and two PG quasars with an order of magnitude higher luminosity \citep{ashton2004} has provided evidence that the distance of the warm absorber from the central engine is proportional to the square root of the ionising luminosity, exactly the same proportionality as the distance of the torus. However, our analysis supports the suggestion \citep{pounds20030844,pounds20031211} that the extremely fast outflows in two nearby quasars originate in accretion disc winds, although we note that their kinetic luminosity, as in the case of the Seyferts, accounts for less than 1\% of the AGN bolometric luminosities. If these outflows are accretion disc winds, then our estimated mass outflow rates and kinetic luminosities provide some benchmarks with which to compare the results of accretion disc models.

We conclude that Seyfert warm absorbers are probably not telling us anything fundamental about the energetics or structure of the central engine, although extreme quasar outflows could potentially be a window into the workings of accretion discs. We find that the amount of matter processed through the AGN system over the lifetime of the outflow can be significant, being potentially the major source of hot ISM in galaxy bulges with 10$^{\rm 7}$ M$_\odot$ black holes, and understanding the influence of this outflow on the host galaxy - and on the evolution of the AGN itself - remains an important challenge.

\begin{acknowledgements}

We wish to thank the anonymous referee for many helpful suggestions, and K. Nandra for useful discussions. AJB and CEA acknowledge the support of PPARC studentships. SVF acknowledges the support of a UK government Overseas Research Students Award and a University College London Graduate School Scholarship. 

\end{acknowledgements}

\end{document}